\documentclass{aa}

\usepackage{microtype}
\usepackage{graphicx}
\usepackage{txfonts}
\begin{document}

   \title{A unifying theory of dark energy and dark matter:\\ Negative masses and matter creation within a modified $\Lambda$CDM framework}
   
   \titlerunning{A unifying theory of dark energy and dark matter}


   \author{J.~S. Farnes
          \inst{1,2} 
          }

   \institute{Oxford e-Research Centre (OeRC), Department of Engineering Science, University of Oxford, Oxford, OX1 3QG, UK.\\
              \email{jamie.farnes@oerc.ox.ac.uk}\thanks{The codes used for the N-body simulations can be downloaded at: https://github.com/jamiefarnes/negative-mass-simulator}
         \and
             Department of Astrophysics/IMAPP, Radboud University, PO Box 9010, NL-6500 GL Nijmegen, the Netherlands.\\
             }

   \date{Received February 23, 2018}

 
\abstract
{Dark energy and dark matter constitute 95\% of the observable Universe. Yet the physical nature of these two phenomena remains a mystery. Einstein suggested a long-forgotten solution: gravitationally repulsive negative masses, which drive cosmic expansion and cannot coalesce into light-emitting structures. However, contemporary cosmological results are derived upon the reasonable assumption that the Universe only contains positive masses. By reconsidering this assumption, I have constructed a toy model which suggests that both dark phenomena can be unified into a single negative mass fluid. The model is a modified $\Lambda$CDM cosmology, and indicates that continuously-created negative masses can resemble the cosmological constant and can flatten the rotation curves of galaxies. The model leads to a cyclic universe with a time-variable Hubble parameter, potentially providing compatibility with the current tension that is emerging in cosmological measurements. In the first three-dimensional N-body simulations of negative mass matter in the scientific literature, this exotic material naturally forms haloes around galaxies that extend to several galactic radii. These haloes are not cuspy. The proposed cosmological model is therefore able to predict the observed distribution of dark matter in galaxies from first principles. The model makes several testable predictions and seems to have the potential to be consistent with observational evidence from distant supernovae, the cosmic microwave background, and galaxy clusters. These findings may imply that negative masses are a real and physical aspect of our Universe, or alternatively may imply the existence of a superseding theory that in some limit can be modelled by effective negative masses. Both cases lead to the surprising conclusion that the compelling puzzle of the dark Universe may have been due to a simple sign error.}


   \maketitle
%

\section{Introduction}\label{intro}
One of the most fascinating aspects of scientific history is that regarding Einstein's efforts with the cosmological constant. It is well known that Einstein added a cosmological constant to his equations in order to provide a static Universe. Due to this bias, he failed to predict the expansion of the Universe that was soon observed by Hubble \citep{1929PNAS...15..168H}. Famously, upon learning of the Universe's expansion, Einstein set the cosmological constant equal to zero and reportedly called its introduction his ``biggest blunder''. 

Most contemporary physicists are familiar with the fact that prior to Hubble's discovery, Einstein associated the cosmological constant term with a constant of integration. However, Einstein did not always believe this to be the case, and important details are currently absent from the historical narrative. In 1918, before famously discarding the cosmological constant, Einstein made the first physical interpretation of the new $\Lambda$ term that he had discovered:
\vspace{-4pt}
\begin{quote}
\emph{``a modification of the theory is required such that `empty space' takes the role of gravitating negative masses which are distributed all over the interstellar space''.} \citep{einstein1918}
\end{quote}
\vspace{-4pt}
Despite this insight, within a year Einstein reformulated his interpretation:
\vspace{-4pt}
\begin{quote}
\emph{``the new formulation has this great advantage, that the quantity $\Lambda$ appears in the fundamental equations as a constant of integration, and no longer as a universal constant peculiar to the fundamental law''}\footnote{The chosen notation used by Einstein was not $\Lambda$, but rather $\lambda$.} \citep{einstein1919}
\end{quote}
\vspace{-4pt}
What led Einstein to believe that negative masses could provide a solution to the cosmological constant is therefore of interest. To understand the physics of negative masses further, we need to `polarise' the Universe so that mass consists of both positive and negative counterparts. Polarisation appears to be a fundamental property of the Universe. Indeed, all well-understood physical forces can be described through division into two opposing polarised states. For example, electric charges (+ and -), magnetic charges (N and S), and even quantum information (0 and 1) all appear to be fundamentally polarised phenomena. It could therefore be perceived as odd that gravitational charges -- conventionally called masses -- appear to only consist of positive monopoles.

While electromagnetism and quantum theory appear quite comprehensively understood, there are numerous indications that our understanding of the nature of mass remains incomplete on all spatial scales. In the standard model of particle physics \citep[e.g.][]{2012PhLB..716....1A}, the mass of fundamental particles such as the nine charged fermions (six quarks and three leptons) and the Higgs boson are all free parameters that cannot be calculated from first principles. In cosmology, the observed matter in the Universe only accounts for 5\% of the observed gravity, while the remaining 26\% and 69\% are accounted for via dark matter and dark energy respectively \citep[e.g.][]{2016A&A...594A..13P}. The physical nature of both these dark phenomena is completely unknown, and the quest to identify the Universe's missing mass has even given rise to modifications to Newton's and Einstein's theories of gravity \citep[e.g.][]{2009Sci...326..812F}. Nevertheless, our understanding of general relativity has been robustly verified by every experimental test \citep[e.g.][]{2016PhRvL.116f1102A}.

Given these substantial challenges to our understanding of mass in the Universe, perhaps a new approach that separates mass into positive and negative polarised counterparts can further our understanding of cosmology. In this case, particles could have the property of positive, zero, or negative mass. In Newtonian physics, this can allow for a variety of different types of negative mass as the inertial and gravitational masses can differ in sign. However, throughout this paper I specifically only consider a negative mass that is consistent with general relativity, so that the weak equivalence principle always holds and negative mass matter always has identical inertial and gravitational mass.

\begin{figure}
   \centering
   \includegraphics[trim=2.3cm 1.618cm 1.0cm 1.0cm,clip=true,angle=0,origin=c,width=8.5cm]{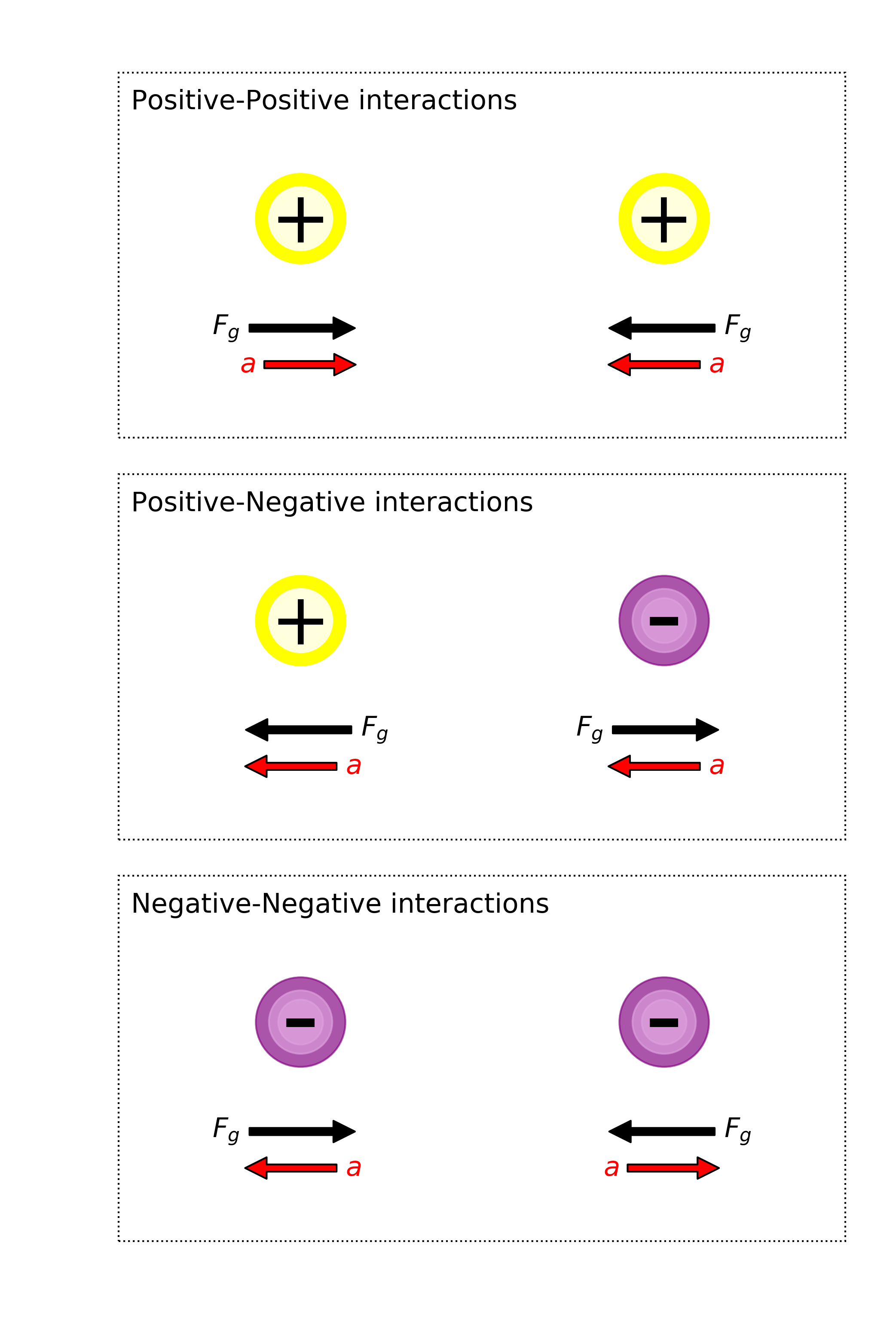}
   \caption{Schematic cartoon of gravitational interactions between positive (in yellow) and negative (in purple) mass particles. Black vectors indicate the direction of the gravitational force, $F_{g} = -G M_{1} M_{2} / r^2$, that is experienced by a given particle. Red vectors indicate the direction of the acceleration, $a=F_{g}/M$, that is experienced by a given particle when the weak equivalence principle holds. There are three possible cases: \textbf{(i) Top row:} the familiar positive--positive mass interaction, in which both particles accelerate towards one another via gravitational attraction, \textbf{(ii) Middle row:} the positive--negative mass interaction, in which both particles accelerate in the same direction -- pointing from the negative mass towards the positive mass. This case is sometimes referred to as runaway motion, and \textbf{(iii) Bottom row:} the negative--negative mass interaction, in which both particles accelerate away from one another via gravitational repulsion.}
              \label{basics}%
    \end{figure}

While positive mass is familiar to all of us, the concept of negative mass is rather exotic.\footnote{A particle with negative mass is not to be confused with a particle of negative mass-squared. Such tachyon particles have an imaginary mass, and are not considered in this paper.} However, such negative masses have a number of basic properties, as shown in Figure~\ref{basics}. While a positive mass gravitationally attracts all surrounding masses, a negative mass will gravitationally repel all surrounding masses. If a force is exerted on a positive mass, the mass will move in the direction of the applied force. However, if a force is exerted on a negative mass, the mass will move towards the applied force. Nevertheless, a negative mass at the surface of the Earth would fall downwards in a similar manner to a positive mass. 

One of the more bizarre properties of negative mass is that which occurs in positive--negative mass particle pairs. If both masses have equal magnitude, then the particles undergo a process of runaway motion. The net mass of the particle pair is equal to zero. Consequently, the pair can eventually accelerate to a speed equal to the speed of light, $c$. Due to the vanishing mass, such motion is strongly subject to Brownian motion from interactions with other particles. In the alternative cases where both masses have unequal magnitudes, then either the positive or the negative mass may outpace the other -- resulting in either a collision or the end of the interaction.
    
Although counterintuitive and ``preposterous'' \citep{bonnor1989}, all of these behaviours violate no known physical laws. Negative masses are consistent with both conservation of momentum and conservation of energy \citep{forward1990}, and have been shown to be fully consistent with general relativity in the seminal work of \citet{bondi1957}. Crucially, negative masses are a natural cold dark matter candidate, as negative mass material could not gravitationally coalesce in order to form astrophysical structures that can initiate fusion and emit light. As negative masses are attracted towards positive masses, they seem capable of applying pressure onto positive masses that could possibly modify a galaxy's rotation curve. Furthermore, negative masses also make a natural dark energy candidate, as a diffuse background of mutually-repelling negative masses could drive the expansion of the Universe. The repulsive negative masses would behave as a dark fluid. The equation of state for a perfect fluid is given by $p=\omega \rho c^2$, where $p$ is the pressure, $\rho$ is the energy density, and $\omega$ is the equation of state parameter\footnote{Sometimes stated in units where $c=1$.}. 

Negative mass is not a new idea and conventional cosmology would suggest that such material is ruled out. Much of the standard reasoning for the exclusion of negative masses is based upon `black swan' arguments, in that a negative mass has never been directly observed. In addition, as the Universe expands, the density of exotic negative mass matter (with $\omega=0$) would be increasingly diluted. The mutually-repulsive nature of negative masses is therefore tantalisingly inconsistent with measurements of dark energy, as this mysterious energy does not dilute but rather retains a constant density (with $\omega=-1$) as the Universe expands \citep[e.g.][]{2014A&A...571A..16P}. This has previously ruled out unusual negative mass matter as a dark energy candidate. The introduction of bimetric models has allowed for extensions of general relativity with two different metrics \citep[e.g.][]{2008PhRvD..78d4015H}. One application of these models has been to explore cosmological theories with negative masses as a form of dark energy \citep{2014Ap&SS.354..611P}, however such theories have remained incompatible with observations.

Despite this initial intellectual hurdle, the basic properties of negative masses still make them a powerful and compelling candidate for providing a unification of dark matter and dark energy within a single theoretical framework. It seems possible that negative masses could constitute a form of dark fluid that can simultaneously explain both of the elusive dark phenomena. Such a simple premise is an elegant one, particularly when considering that the introduction of negative mass adds symmetry to the Universe. Nevertheless, a theory of negative mass is complicated to assess in relation to other known cosmological results -- which have overwhelmingly been developed with the implicit assumption that mass is only positive. For these reasons, nearly all theories, experiments, observations, and physical interpretations need to be rigorously revisited. However, if negative mass is a conceptual blindspot that is stopping us from developing a more advanced theory of the Universe, then such elaborations can clearly be fruitful. In this paper, I therefore attempt to test whether Einstein's insight could be correct, and what the consequent implications would be for modern cosmology. 

The paper is structured as follows: in Section~\ref{background}, I modify general relativity to consider the theory of a negative mass cosmology. In Section~\ref{numericalresults}, I present three-dimensional N-body simulations of positive and negative masses using 50,000 particles. In Section~\ref{compatibility}, I consider any potential compatibility between cosmological observations of distant supernovae, the cosmic microwave background, and galaxy clusters with the negative mass hypothesis. Section~\ref{discussion} provides a speculative discussion on outstanding theoretical considerations that can be explored in the future, and in Section~\ref{summary} I provide a summary of the results.

\section{Theoretical results}
\label{background}
To understand the cosmological implications of negative masses, I begin with Einstein's field equations,
\begin{equation}
R_{\mu \nu} - \frac{1}{2} R g_{\mu \nu} + \Lambda g_{\mu \nu} = \frac{8 \pi G}{c^4} T_{\mu \nu} ,
\label{EFE}
\end{equation}
where $R_{\mu \nu}$ is the Ricci curvature tensor, $R$ is the scalar curvature, $g_{\mu \nu}$ is the metric tensor, $\Lambda$ is the cosmological constant, $G$ is Newton's gravitational constant, $c$ is the speed of light in vacuum, and $T_{\mu \nu}$ is the stress-energy tensor.

Assuming a homogeneous and isotopic universe via the Friedmann--Lema\^{i}tre--Robertson--Walker (FLRW) metric, and assuming that the stress-energy tensor is that of a perfect fluid, Eqs.~(\ref{EFE}) can be used to derive the Friedmann equation,
\begin{equation}
\left( \frac{\dot{a}}{a} \right)^2 = \frac{8 \pi G}{3}\rho + \frac{\Lambda c^2}{3} - \frac{k c^2}{a^2} ,
\label{friedmann}
\end{equation}
and the Friedmann acceleration equation,
\begin{equation}
\frac{\ddot{a}}{a} = -\frac{4 \pi G}{3} \left(\rho + \frac{3 p}{c^2} \right) + \frac{\Lambda c^2}{3},
\label{friedmannaccn}
\end{equation}
where $a$ is the scale factor, $H \equiv \dot{a}/a$ is the Hubble parameter, $\rho$ is the total mass density of the universe, $p$ is the pressure, $k$ is the curvature parameter or intrinsic curvature of space, and $k/a^2$ is the spatial curvature in any time-slice of the universe. $k=+1$, $0$, and $-1$, indicate a closed, flat, and open universe respectively. For a perfect fluid, the pressure and mass density of the fluid are related via the equation of state, which is given by $p=\omega \rho$, where $\omega$ is some constant and for negative mass matter $\omega=0$, this means that for negative mass matter the pressure is negligible relative to the mass density. However, I will show that $\omega$ in the equation of state can be modified in cases of matter creation or annihilation. For gravitationally repulsive matter (such as negative masses) that is constantly being created, $\omega$ can equal $-1$ as described in Section~\ref{mattercreation-section}.

For a cosmology that contains negative masses, I can also separate the content of the universe into its respective fractional contributions from the density parameters of the various species of energy contributors, such that,
\begin{equation}
\Omega_{M+}+\Omega_{M-}+\Omega_{r}+\Omega_{\Lambda}=\Omega ,
\end{equation}
where $\Omega=1-\Omega_{k}$ and the density contributions $\Omega_{i}$ include positive masses $M+$, negative masses $M-$, radiation $r$, curvature $k$, and the cosmological constant $\Lambda$. By definition, $\Omega_{M-}$ is necessarily negative or zero-valued. In the conventional $\Lambda$CDM cosmology \citep[e.g.][]{2014A&A...571A..16P,2016A&A...594A..13P}, it is implicitly assumed that $\Omega_{M-}=0$ and explicitly assumed that $\Omega_{\Lambda}>0$. Moreover, $\Omega_{M+}$ (usually given the standard notation $\Omega_{M}$) is typically separated into additional constituents, such as $\Omega_{b}$ for baryons and $\Omega_{\textrm{CDM}}$ for cold dark matter. To date, no attempt has been made to discern negative mass contributions from observations of the cosmic microwave background (CMB) \citep[e.g.][]{2016A&A...594A..14P}.

In the case of a matter-dominated universe (which has $\Omega_{r}=\Omega_{\Lambda}=0$) that includes negative masses, there are three possibilities: (i) a positive-mass dominated cosmology with $|\Omega_{M+}|>|\Omega_{M-}|$, (ii) a massless cosmology with $|\Omega_{M+}|=|\Omega_{M-}|$, and (iii) a negative-mass dominated cosmology with $|\Omega_{M+}|<|\Omega_{M-}|$. Eqs.~(\ref{EFE}) to (\ref{friedmannaccn}) can therefore be solved for each of these three cases. 


\subsection{Positive mass cosmology}
\label{positivemasses}
The positive mass dominated cosmology with $|\Omega_{M+}|>|\Omega_{M-}|$ corresponds to the standard matter-dominated universe solutions with a critical density given by $\rho_{c}=3H^2 / 8 \pi G$ and a total density parameter given by $\Omega=\rho/\rho_c$, where $\Omega=1$, $<1$, and $>1$, correspond to critical density, open, and closed universes respectively \citep[e.g.][]{1924ZPhy...21..326F}.

The key difference to a conventional matter-dominated cosmology is that the Friedmann equation is modified to,
\begin{equation}
\left( \frac{\dot{a}}{a} \right)^2 = \frac{8 \pi G}{3} \left(\rho_{+} + \rho_{-} \right) ,
\label{friedmann2}
\end{equation}
where $|\rho_{+}| > |\rho_{-}|$ and $\Omega=(\rho_{+}+\rho_{-})/\rho_c$. In this new context the meaning of the critical density is not clear. Typically, the critical density is the specific value of positive mass density at which the universe will not expand forever and will also not recollapse. This meaning remains broadly unchanged for a negative mass cosmology, with the sole caveat that $\rho_c$ now becomes the net density required to prevent both eternal expansion and recollapse. This can be illustrated by observing that a critical density universe corresponds to $(\rho_{+}+\rho_{-})=\rho_{c}$.


\subsection{Massless cosmology}
\label{massless-cosmo}
The massless cosmology with $|\Omega_{M+}|=|\Omega_{M-}|$ is equivalent to an empty universe. Observations clearly indicate that the Universe is not empty. However, if the Universe contains equal quantities of positive and negative masses then this is not an unphysical solution -- the Universe can have zero mass on large-scales. This is equivalent to setting $T_{\mu \nu}=0$ in Eqs.~(\ref{EFE}), $\rho=0$ in Eqs.~(\ref{friedmann}) to (\ref{friedmannaccn}), and $\Omega=0$. The consequences have been previously studied both directly and indirectly.

\citet{1973Natur.246..396T} demonstrated that sufficient negative mass energy in the Universe would naturally allow for the creation of a cosmos from nothing -- with the Big Bang then being an energy-conserving event. The Universe could then be a zero mass, zero energy, system. In quantum field theory, every event that could happen in principle actually does happen occasionally in a probabilistic manner. This allows for the spontaneous and temporary emergence of particles from the vacuum in the form of vacuum fluctuations. In essence, particles can borrow energy from the vacuum to come into existence for a brief period, as long as that energy is repaid. This process can be described through the uncertainty principle, $\Delta E \Delta t \sim \hbar/2$, which leads to the natural consequence that our Universe could simply be a vacuum fluctuation that has zero energy and can therefore exist eternally. This implies that our Universe is just one of those things that happen on occasion, and we can simply think of its existence as being illustrated by a 1~billion-$\sigma$ statistical event. In \citet{1973Natur.246..396T}, this negative energy was suggested to be the gravitational potential energy from a positive mass due to its interaction with the rest of the Universe, although a negative mass energy can achieve the same function.

A massless cosmology with negative masses has also been considered before \citep{2012A&A...537A..78B}. This cosmology has symmetric quantities of matter and antimatter, in which antimatter is hypothesised to have negative gravitational mass, thereby leading to gravitational repulsion between matter and antimatter. The cancellation of positive and negative masses yields an effectively empty space-time known as the Dirac--Milne universe. In this cosmology, the scale factor evolves linearly with time, thereby directly solving the age and horizon problems of a matter-dominated universe so that inflation is no longer required. While this is an elegant solution, such a cosmology would violate the weak equivalence principle and therefore appears to be inconsistent with general relativity \citep{bondi1957}. Furthermore, if negative mass particles were antimatter, one would expect positive--negative mass collisions to give rise to a diffuse gamma ray background, which is not observed. Some of the earliest descriptions of antimatter did consider those particles to have a negative mass, and for collisions to result in a release of energy \citep[e.g.][]{schuster1898}. However, it has also been suggested that the collision of negative and positive masses would result in nullification \citep{forward1990}. 

These former works have demonstrated that the introduction of negative mass to cosmology has the capacity to completely cancel the energy budget of the Universe. Nevertheless, this precise cancellation is not a necessity, and is rather a special case. A matter-dominated universe could predominantly consist of positive masses (as for the familiar case presented in Section~\ref{positivemasses}), or could be dominated by negative masses.


\subsection{Negative mass cosmology}
\label{neg-cosmo}
The negative mass dominated cosmology with $|\Omega_{M+}|<|\Omega_{M-}|$ corresponds to a universe that is less than empty. This is equivalent to setting $T_{\mu \nu}<0$ or $\rho<0$. In a non-expanding space, a negative mass fluid would undergo an expansion that is characterised by a uniform stretching as detailed in Appendix~\ref{appendix-a}. However, in this Section we are concerned with the cosmological evolution of an expanding space.

For a matter-dominated cosmology with $\Lambda=0$, purely via inspection of Eq.~(\ref{friedmann}), it is clear that if $\rho$ is negative, there can only be physical solutions when $k=-1$. The negative mass cosmology therefore has negatively curved spatial sections. We can rewrite the Friedmann equation as
\begin{equation}
\left( \frac{\dot{a}}{a} \right)^2 = H^2 = \frac{8 \pi G}{3}\rho_{-} + \frac{c^2}{a^2} .
\label{friedmann3}
\end{equation}

For a matter-dominated cosmology with a cosmological constant, if $\rho$ is negative and $\Lambda>0$, it can be possible that $k=-1, 0, +1$. In the alternative case, where $\rho$ is negative and $\Lambda<0$, there can again only be physical solutions when $k=-1$. In both cases
\begin{equation}
\Omega_{M+}+(\Omega_{M-}+\Omega_{\Lambda})=1-\Omega_{k} ,
\end{equation}
and so it can be seen that a degeneracy exists between $\Omega_{M-}$ and $\Omega_{\Lambda}$ such that $\Omega_{\mathrm{degen}}=\Omega_{M-}+\Omega_{\Lambda}$. In this case, a degeneracy is meant in the sense that one can change both parameters and leave their sum unchanged. In a conventional $\Lambda$CDM cosmology, where $\Omega_{M-}$ is taken to be zero, one is actually measuring $\Omega_{\mathrm{degen}}$ and could therefore falsely infer a cosmological constant instead of a negative density parameter. 

It is typically assumed that $\Omega_{M-}$ and $\Omega_{\Lambda}$ can be discriminated based upon the equation of state for a perfect fluid. For non-relativistic matter, $\omega=0$, whereas for a true cosmological constant, $\omega=-1$. In other words, negative mass matter is conventionally assumed to dilute in density as the Universe expands, whereas the cosmological constant remains constant. From observations, the equation of state parameter for the $\Omega_{\mathrm{degen}}$ component is close to $-1$ \citep[e.g.][]{2014A&A...571A..16P}. This is typically interpreted as evidence for mysterious dark energy, the physical properties of which remain unknown.

\subsubsection{Matter creation}
\label{mattercreation-section}
I have demonstrated that the degeneracy between negative mass matter and a cosmological constant can typically be broken using the equation of state parameter, $\omega$. However, I now consider the consequences if matter is continuously created in the Universe. Such matter creation is not a new concept and was previously invoked to describe the disproven steady-state theory of the Universe \citep{1948MNRAS.108..252B,1960MNRAS.120..256H}. In this theory, new matter is continuously-created in order to keep the average density constant as particles move apart due to the expansion of the Universe. This steady-state theory is inconsistent with observations, as it cannot explain the cosmic microwave background or radio source counts. However, the creation term has only ever been applied to positive mass matter, and so may still be useful for understanding dark energy and matter in an expanding Universe. 

Matter creation has been explored more recently in terms of particle production at the expense of a gravitational field \citep{1988PNAS...85.7428P}. I now calculate the implications for negative masses. One plausible approach is to model the cosmological solutions for `adiabatic' particle production in the presence of spatial curvature or an equivalent `fluid'. For creation of matter, we can modify Einstein's equations to include a creation term or a C-field to the effective energy-momentum tensor,
\begin{equation}
T_{\mu \nu}^{\prime} = T_{\mu \nu} + C_{\mu \nu},
\label{creation}
\end{equation}
where $T_{\mu \nu}$ is the energy-momentum tensor for a fluid with equation of state $p=(\gamma-1)\rho$ such that
\begin{equation}
T_{\mu \nu} = (\rho+p) u_{\mu} u_{\nu} + p g_{\mu \nu} ,
\label{creation-T}
\end{equation}
and $C_{\mu \nu}$ is the energy-momentum tensor which corresponds to the matter creation term, such that
\begin{equation}
C_{\mu \nu} = P_{C} (g_{\mu \nu} + u_{\mu} u_{\nu}) .
\label{creation-C}
\end{equation}
Einstein's field equations are therefore modified to
\begin{equation}
R_{\mu \nu} - \frac{1}{2} R g_{\mu \nu} + \Lambda g_{\mu \nu} = \frac{8 \pi G}{c^4} \left( T_{\mu \nu} + C_{\mu \nu} \right) ,
\label{EFE-mattercreation}
\end{equation}
and the two fluids interact to provide an effective equation of state $p=\omega_{\mathrm{eff}} \rho$ \citep[e.g.][]{2016MNRAS.460.1445P}. The resulting field equations do not just describe space and matter, but now also include creation. The creation tensor cancels the matter density and pressure, leaving just the overall effective form of the vacuum tensor. The sources of the creation tensor are places where matter is created (or destroyed). The $C_{\mu \nu}$ term has previously been used to suggest an alternative theory of gravity \citep{1964RSPSA.282..178H}, however such a field was suggested to be incompatible with an expanding universe unless matter with negative inertial and gravitational mass were to exist \citep{1965RSPSA.286..313H}.

For a FLRW metric in which matter creation is taking place, Eq.~(\ref{friedmannaccn}) is modified to include an additional term due to creation pressure, which is related to gravitationally induced adiabatic particle production by
\begin{equation}
P_{C} = \frac{-\Gamma}{3H} (p+\rho) ,
\label{creationpressure}
\end{equation}
where $P_{C}$ is the matter creation pressure, $\Gamma(t)$ is the creation rate or more specifically the rate of change of the particle number in a physical volume $V$ containing $N$ particles \citep{2016MNRAS.460.1445P,2017PhRvD..95j3516P}. By assuming that the fluid component of Eqs.~\ref{friedmann} and \ref{friedmannaccn} behaves as a perfect fluid with an equation of state given by $p=(\gamma-1)\rho$, then \citet{2017PhRvD..95j3516P} show that the relation between $\omega_{\mathrm{eff}}$ and the particle creation rate $\Gamma$ is given by
\begin{equation}
\omega_{\mathrm{eff}} = \frac{P}{\rho} = -1 + \gamma \left( 1 - \frac{\Gamma}{3H} \right) ,
\label{effectivestate}
\end{equation}
where $P=p+P_{C}$, and $\gamma$ is a constant. This has an important consequence as it shows that during matter creation, the equation of state parameter is related to the particle creation rate. Hence, different effective fluids and different gravitational effects can be provided by varying $\Gamma$. In other words, the creation of matter modifies the equation of state parameter. For a constant creation rate of $\Gamma=3H$, one obtains $\omega=-1$. The introduction of matter creation can therefore yield an equation of state equivalent to a cosmological constant.

It is worth noting that the equation of state parameter of this space-time fluid, $\omega_{\textrm{eff}}$, can actually take on many values depending on the precise rate of matter creation, and it is only constrained to exactly equal $-1$ when $\Gamma=3H$. Only when the creation of negative masses sustains itself at a very specific rate does it serve to provide a constant density as a function of time and thus a cosmological constant with $\omega_{\textrm{eff}}=-1$. If the creation of negative masses is occurring too slowly to replenish the density, which is steadily decreasing through mutual repulsion and the expansion of the universe, then $\omega_{\textrm{eff}}>-1$. If the density of negative masses is increasing as a function of time, then we obtain a phantom energy model with $\omega_{\textrm{eff}}<-1$. Further study of the creation term $C_{\mu \nu}$ will be able to determine what drives the creation rate. However, within this model the creation rate $\Gamma$ can vary as both a function of space and time. When varying as a function of time, $\Gamma(t)$ potentially gives rise to a time-varying Hubble parameter and a time-varying cosmological constant. When varying as a function of space, this gives rise to an inhomogeneous distribution of $\omega_{\textrm{eff}}$ throughout the universe. For example, one could imagine spatial pockets of negative mass with particularly high or low creation rates. This would give rise to an inhomogeneous and anisotropic distribution of expansion speeds across the sky that varies around individual voids in the galaxy distribution, and which could therefore be tested using observational data. This can be done by looking at the measured galaxy distribution using upcoming radio telescopes such as the Square Kilometre Array (SKA) and its pathfinders and precursors in combination with future optical and infrared surveys such as Euclid and LSST \citep[e.g.][]{2015arXiv150104076M}. Future studies will be able to determine whether there is matter creation taking place and to investigate in more detail any influence that $\Gamma$ could have on the ultimate fate of the Universe.

The continuous-creation of negative masses implies that the universe would be taking on an increasingly negative energy state. I note that this negative energy is not created all at once in a cataclysmic event, but is steadily created at the rate given by $\Gamma$. This could be interpreted as the vacuum instability being a real, physical, phenomena. In other words, rather than positive mass matter collapsing into an infinitely negative energy state, this cosmology results in the continuous creation of (potentially real or virtual) particles with negative energy. Such speculations can be considered more rigorously in future works.

\subsubsection{Equivalence with the cosmological constant}
\label{equivalence}
Let us now consider the case of negative masses that are being constantly created. These negative masses gravitationally repel each other -- thereby pushing apart structures in the universe. I showed in Section~\ref{mattercreation-section} that a constant creation rate of $\Gamma=3H$ provides $\omega=-1$. Negative masses that are being constantly created therefore do not undergo the typical dilution as the universe expands, as would be the case for normal matter which conventionally has $\omega=0$. In fact, these continuously-created negative masses appear to resemble dark energy.

There is therefore a strong degeneracy between $\Omega_{M-}$ and $\Omega_{\Lambda}$, as discussed in Section~\ref{neg-cosmo}. This degeneracy is present in the supernovae observations of an accelerating Universe (see Section~\ref{compatibility}).

I am arguing that these negative masses are created at such a rate that they retain -- on large scales -- a constant density. I can therefore modify Friedmann's equation to
\begin{equation}
\left( \frac{\dot{a}}{a} \right)^2 = \frac{8 \pi G}{3} \left( \rho_{+} +  \rho_{-} \right) - \frac{k c^2}{a^2} .
\label{friedmann-mod}
\end{equation}
As $\rho_{-}$ is constant, I can define $\Lambda=8\pi G \rho_{-} /c^2$, thereby obtaining
\begin{equation}
\left( \frac{\dot{a}}{a} \right)^2 = \frac{8 \pi G}{3}\rho_{+} + \frac{\Lambda c^2}{3} - \frac{k c^2}{a^2} .
\label{friedmann-final}
\end{equation}
This is the standard Friedmann equation given in Eq.~(\ref{friedmann}). In conclusion, I have shown that constantly-created negative masses are a natural explanation for the cosmological constant. The physical properties of negative masses that were undergoing constant creation would be rather similar to the accepted and inferred properties of the cosmological constant. Within this toy model, $\Lambda$ can now be interpreted as the neglected negative mass content of the Universe, rather than as some mysterious extra term. As these negative masses can take the form of a cosmological constant, one can deduce that they are a property of the vacuum rather than non-relativistic matter in the conventional sense. In summary, the field equations are modified to
\begin{equation}
R_{\mu \nu} - \frac{1}{2} R g_{\mu \nu} = \frac{8 \pi G}{c^4} \left( T_{\mu \nu}^{+} + T_{\mu \nu}^{-} + C_{\mu \nu} \right) ,
\label{EFE-modified}
\end{equation}
where the conventional $\Lambda g_{\mu \nu}$ term is now represented by a combination of $T_{\mu \nu}^{-}$ (an exotic matter term) and $C_{\mu \nu}$ (a modified gravity term).

\subsubsection{Solutions to the Friedmann equation}
\label{AdS-space}
I showed in Section~\ref{equivalence} that $\Lambda$ can be modelled as a constant density of negative masses, which remain constant via matter creation. If I consider the positive mass matter as negligible, we have the Friedmann equation
\begin{equation}
\left( \frac{\dot{a}}{a} \right)^2 = \frac{\Lambda c^2}{3} - \frac{k c^2}{a^2} ,
\label{friedmann-ads}
\end{equation}
where $\Lambda=8\pi G \rho_{-} /c^2$. As by definition $\rho_{-}<0$, it must also be the case that $\Lambda<0$. It is clear that for $\Lambda<0$, Eq.~(\ref{friedmann-ads}) can only be satisfied when $k=-1$. The solution for the scale-factor is given by
\begin{equation}
a(t) = \sqrt{\frac{-3}{\Lambda c^2}} \sin \left( \sqrt{\frac{-\Lambda c^2}{3}} t \right) ,
\label{friedmann-soln}
\end{equation}
together with a Hubble parameter given by
\begin{equation}
H(t) = \sqrt{\frac{-\Lambda c^2}{3}} \cot \left( \sqrt{\frac{-\Lambda c^2}{3}} t \right) ,
\label{hubble-soln}
\end{equation}
which corresponds to an Anti-de Sitter space (AdS) with a time-variable Hubble parameter. This AdS universe undergoes a cycle of expansion and contraction with a timescale of $\sqrt{-3\pi^2/\Lambda c^2}$. This is as a negative value of $\Lambda$, if interpreted as vacuum energy, corresponds to a negative energy density \citep{1982CMaPh..87..577H}. This is a counterintuitive result, as although the negative masses are gravitationally repelling one another, the cosmological effect appears to be for the negative energy associated with negative masses to cause the universe to recollapse. The solution describes an open universe which expands from a Big Bang, reaches a maxima, and then recontracts to a Big Crunch. This appears to be an elegant outcome -- by introducing negative masses that undergo continuous creation, one obtains a cyclic cosmology. Even with the addition of positive mass matter, a universe with a negative cosmological constant would eventually recollapse due to this extra attractive force. Interestingly, the value of $\Lambda$ is necessarily finely-tuned -- if the magnitude of $\Lambda$ is too large, then the universe will not exist for sufficient time to create observers. For a cosmological constant of similar magnitude to that which is currently interpreted from cosmological observations, $|\Lambda| \approx 0.3\times10^{-52}$~m$^{-2}$, the lifetime of each cycle of the universe would be $\sim105$~Gyr. In an epoch $\sim14$~Gyr after a Big Bang, the universe would therefore be observed to be in an expanding phase. The continuous negative mass matter creation would cause objects in this universe to be gravitationally accelerated apart, but only on a local level. In late-times, this cyclic cosmology with AdS space collapses due to its negative energy.

One could also attempt to calculate the age of the universe in this cosmology, which can be conveniently estimated from Eq.~\ref{hubble-soln} as
\begin{equation}
t_{0} = \sqrt{\frac{-3}{\Lambda c^2}} \arctan \left( \sqrt{\frac{-\Lambda c^2}{3}} \frac{1}{H_{0}} \right) .
\label{age-soln}
\end{equation}
For $H_{0} = 67$~km~s$^{-1}$~Mpc$^{-1} = 2.17\times10^{-18}$~s$^{-1}$ and $|\Lambda| \approx 0.3\times10^{-52}$~m$^{-2}$, then the age of the universe would be $\sim13.8$~Gyr. While this appears to be compatible with observations of our own Universe, there are several notable caveats that could either extend or shorten the calculated age. In the presented toy model, $\Lambda$ can itself be a function of time, which could drastically alter the results. The fundamental change of including negative masses into the Universe would also require reinterpretation of the standard observational results -- which may in turn adjust the values of both $H_{0}$ and $\Lambda$.

One notable point of interest is that the derived AdS space corresponds to one of the most researched areas of string theory, the Anti-de Sitter/Conformal Field Theory correspondence \citep{1999IJTP...38.1113M}. String theories naturally require the compactification of six dimensional Calabi-Yau spaces onto AdS backgrounds, meaning onto space-times with negative $\Lambda$ \citep[e.g.][]{2011arXiv1105.0078P}, and it has therefore remained puzzling why cosmological observations have not provided evidence for $\Lambda<0$. In the proposed toy model, $\Lambda<0$ also seems to be able to explain the flat rotation curves observed in galaxies, as detailed in Section~\ref{flatten-theory}. The existence of AdS space implies that string theory may possibly be directly applicable to our Universe. I note that this cyclic universe does not violate the positive energy theorem of \citet{witten1981}, which approximately states that it is not possible to construct an object out of ordinary matter (with positive local energy density), that has a total energy which is negative. The positive energy theorem has also been extended to AdS space-time \citep{abbott1982,gibbons1982}. 

Another interesting result is that the theory directly predicts a time-variable Hubble parameter, which may be consistent with recent cosmological measurements \citep{2014MNRAS.440.1138E,2016ApJ...826...56R,2017MNRAS.465.4914B}. Nevertheless, in this scenario, estimates of $H$ from the CMB would likely be strongly affected through modelling the CMB data with a flat $\Lambda$CDM model universe. I note that I have here only considered the effects upon a cosmology dominated by $\Lambda$ and $k$. Although beyond the scope of this current paper, consideration of a more physical universe -- that also includes $\rho_{+}$ -- may considerably adjust the cosmological behaviour at early times and the precise form of a time-variable Hubble parameter. Such scenarios can be explored further in future works.

\subsubsection{Flattening of galaxy rotation curves}
\label{flatten-theory}
I now consider the localised properties of positive masses that are immersed in a negative mass fluid. Such a scenario will clearly have an effect on the dynamics of positive masses and the subsequent evolution of a positive mass system. This thereby allows us to consider the implications for the dynamics of galaxies and similar structures. 

The standard galaxy rotation scenario can be described as the case where a positive mass test particle, of mass $M_{+}$, is located at a distance $r$ from the galactic centre. The galaxy has the majority of its mass, $M_{\star}$, concentrated at the centre. I consider this system from an observing frame where the positive mass particle is seen to be rotating with an orbital velocity, $v$, about an axis of rotation through the centre of the galaxy. For a stable circular orbit, the gravitational force (which acts inwardly towards the orbital centre) is equal to the centripetal force (which is related to the component of the velocity acting tangentially to the orbital path). This provides the simple equation
\begin{equation}
\frac{G M_{+} M_{\star}}{r^2} = \frac{M_{+} v^2}{r} .
\label{DM-one}
\end{equation}
By rearranging Eq.~(\ref{DM-one}), I obtain
\begin{equation}
v = \sqrt{\frac{G M_{\star}}{r}} ,
\label{DM-two}
\end{equation}
which is the standard Keplerian rotation curve whereby the orbital velocity of a test particle should decrease as a function of increased radius. Nevertheless, very few galaxies show any evidence for such a Keplerian rotation curve. H$\alpha$ and radio HI observations of galaxies both indicate that rotation curves remain essentially flat out to several tens of kpc, providing strong evidence for the existence of dark matter \citep{1970ApJ...159..379R,1980ApJ...238..471R,1985ApJ...289...81R}.

I now construct an alternative model for the galaxy rotation scenario, that also includes a cosmological constant. In this case, the same setup exists as in the standard galaxy rotation scenario. However, I also now assume that the positive mass galaxy is surrounded by a halo of continuously-created negative masses, with constant density $\rho_{-}$ and of total mass $M_{-}$. The positive mass test particle is now immersed in a negative mass fluid that, as shown in Section~\ref{equivalence}, behaves with resemblance to a cosmological constant with $\Lambda=8\pi G \rho_{-} /c^2$.
 \begin{figure}
   \centering
   \includegraphics[clip=false,angle=0,origin=c,width=\hsize]{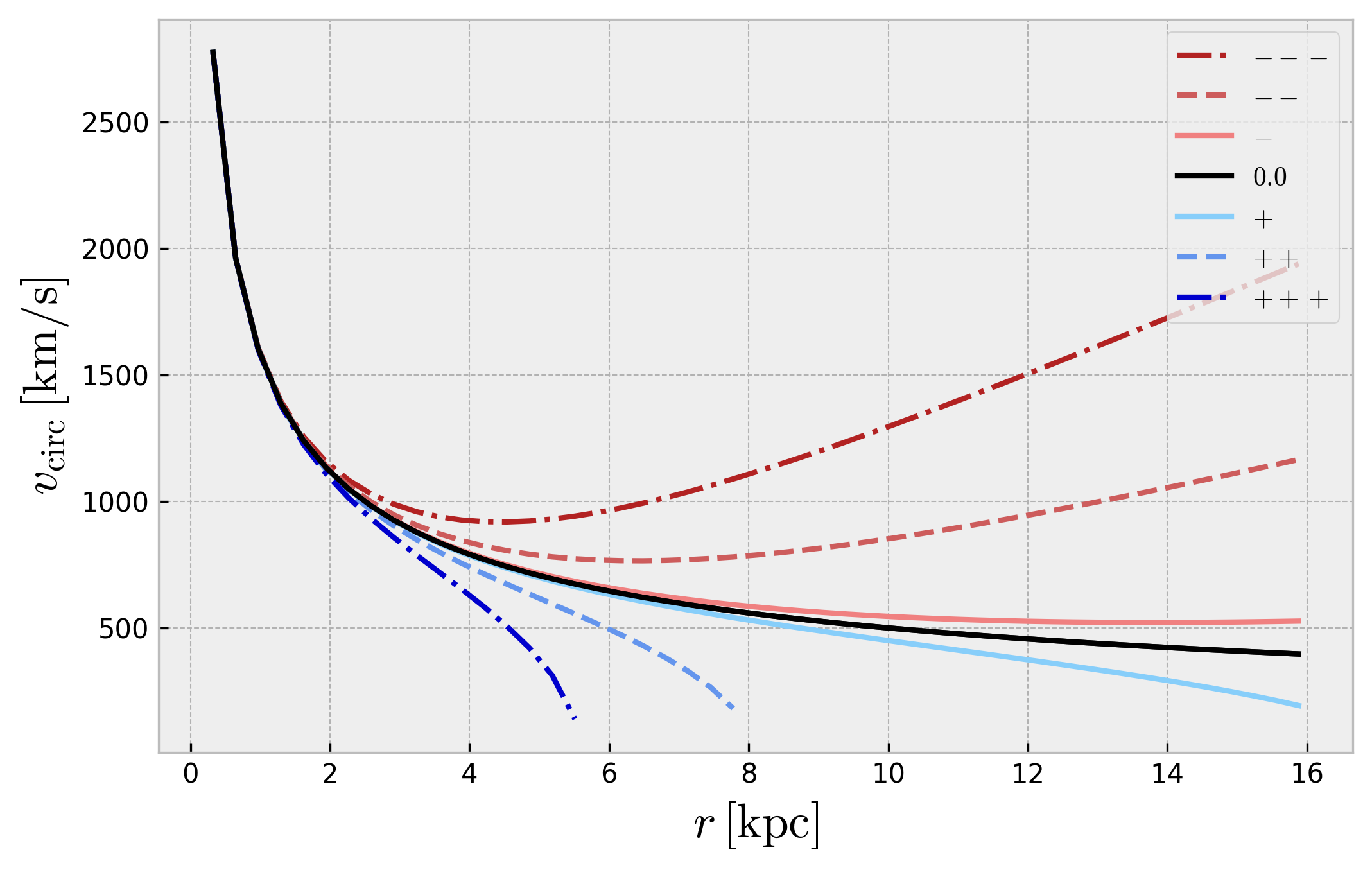}
   \caption{Predicted circular velocity as a function of radius, for a galaxy of similar size and mass to the Milky Way and that is influenced by a cosmological constant. The displayed rotation curves are for increasing magnitudes of a positive (in blue) and a negative (in red) cosmological constant. The Keplerian curve with $\Lambda=0$ is also shown (in black). Whereas a positive cosmological constant steepens the decline of a rotation curve, a negative cosmological constant flattens the rotation curve, causing a steady increase at larger galactic radii. Solid body rotation in the centre of the galaxy is not shown.}
              \label{rotationcurve-sim}%
    \end{figure}

In the limit of small velocity, $v \ll c$, and in the weak field approximation, Einstein's field equations should reproduce Newtonian gravity. We can therefore examine the Newtonian limit of the field equations in the case of a non-zero cosmological constant. We can define
\begin{equation}
\rho_{\textrm{vac}} = \frac{\Lambda c^2}{4 \pi G} ,
\label{DM-3}
\end{equation}
so that the Poisson equation for gravity is given by
\begin{equation}
\nabla^2\phi = 4 \pi G \rho_{+} - 4 \pi G \rho_{\textrm{vac}} = 4 \pi G \rho_{+} - \Lambda c^2 ,
\label{DM-4}
\end{equation}
where $\phi$ is the scalar potential and $\rho_{+}$ is the positive mass density. If we set $\phi=0$ at the origin and assume spherical symmetry, then Eq.~(\ref{DM-4}) has the solution
\begin{equation}
\phi(r) = -\frac{G M_{\star}}{r} -\frac{\Lambda c^2}{6} r^2 ,
\label{DM-5}
\end{equation}
which gives the potential at distance $r$ from a central point mass $M_{\star}$. In the case where $\Lambda=0$, we retrieve the standard potential for Newton's law of gravitation. The force exerted on a particle of mass $m$ is related to the scalar potential by $F=-m\nabla\phi$, which yields
\begin{equation}
F = -m \frac{\partial}{\partial r}\phi = -m \frac{\partial}{\partial r} \left[ -\frac{G M_{\star}}{r} -\frac{\Lambda c^2}{6} r^2 \right] = -\frac{G M_{\star} m}{r^2} +\frac{\Lambda c^2}{3} m r .
\label{DM-6}
\end{equation}
We can therefore modify Eq.~(\ref{DM-one}) to provide
\begin{equation}
\frac{m v^2}{r} = \frac{G M_{\star} m}{r^2} -\frac{\Lambda c^2}{3} m r ,
\label{DM-7}
\end{equation}
which by rearrangement, leads to the solution for the rotation curve
\begin{equation}
v = \sqrt{\frac{G M_{\star}}{r} -\frac{\Lambda c^2}{3} r^2} .
\label{DM-8}
\end{equation}
Again, we can see that for $\Lambda=0$, we obtain the standard Keplerian rotation curve. However, for non-zero values of the cosmological constant, the rotation curve is modified. The modification of the rotation curve for increasing magnitudes of positive and negative $\Lambda$ in a galaxy with mass and size similar to the Milky Way is shown in Figure~\ref{rotationcurve-sim}. For a negative cosmological constant, the rotation curve clearly increases linearly towards the outskirts of a galaxy, such that $v \propto (|\Lambda|^{1/2} c / \sqrt{3}) r$. This appears consistent with observational results, and previous studies have found that most rotation curves are rising slowly even at the farthest measured point \citep{1980ApJ...238..471R}. The data again appear to be consistent with a negative cosmological constant. I emphasise that the rotation curves are being affected by the local negative mass density, which can coalesce into halo-like structures (see Section~\ref{DMhalo-sim}). The inferred value of $\Lambda$ in the vicinity of galaxies is therefore expected to be of larger magnitude than the cosmological constant inferred from cosmological expansion. In other words, although a specific large-scale constant rate of particle creation is required to enable a $\omega_{\mathrm{eff}} = -1$ universe, the resulting particle density in intergalactic space would likely be less than the amount of negative masses that would be required to balance a galaxy rotation curve. Hence, more generally the solution for the rotation curve is given by
\begin{equation}
v = \sqrt{\frac{G M_{\star}}{r} -\frac{8 \pi G \rho_{-}}{3} r^2} ,
\label{DM-general}
\end{equation}
where $\rho_{-}$ is the local negative mass density.

I note that the presented solutions, which require a steeply-rising linear rotation curve, are only applicable for a strongly negative $\Lambda$. Figure~\ref{rotationcurve-sim} demonstrates that for a slightly negative $\Lambda$, the increase in velocity is just able to overcome the Keplerian decline, such that the rotation curve remains essentially flat as a function of distance.

Although the majority of observed rotation curves are largely flat, there is some observational evidence that rotation curves can continue to rise out to large galactocentric radii \citep[e.g.][]{1980ApJ...238..471R}. However, there are also reasons why the theoretical rotation curves may generally appear to rise more rapidly than is often seen in observations. The presented toy model is just a simple case, which demonstrates the change to a rotation curve enacted by a negative mass halo. There are several simplifying assumptions that have been made: (i) the positive mass within a galaxy is not concentrated at a central point and has a radially-dependent mass profile, (ii) a typical galaxy has separate bulge, halo, and disk components which are not modelled here, and (iii) the local negative mass density may be asymmetric and itself has a radial mass distribution. In order to fit observational rotation curve data, these additional factors would all need to be modelled in further detail. This would then be able to either validate or rule out the proposed cosmology (and the existence of a negative cosmological constant). However, we consider this fitting process to be beyond the scope of this current paper.

Interestingly, a solution that invoked a negative cosmological constant as an explanation for flat galaxy rotation curves has been considered before by \citet{1999astro.ph.11485W,2001sddm.symp...66K}, but was dismissed by \citet{2000astro.ph..2152B} on the basis that it appeared incompatible with observations of supernovae. In Section~\ref{compatibility}, a proposal is made to provide reconciliation between a negative cosmological constant and the supernovae results.


\section{N-body simulations}
\label{numericalresults}
One of the most effective manners of testing a physical hypothesis related to particle interactions is via N-body simulations. Most modern N-body software packages do not support exotic and rarely-studied phenomena such as negative masses. I have therefore written new software to perform three-dimensional (3D) gravitational N-body simulations using \textsc{python}, \textsc{numpy}, and \textsc{matplotlib}. The code is parallelised using \textsc{dask} in order to make use of the multiple processing cores available in most modern machines. The N-body code used here evaluates the particle positions and velocities at each timestep by using direct methods, thereby avoiding the introduction of any approximations and maintaining the highest accuracy -- albeit at the cost of substantial computing time, of the order of $O(N^2)$ per timestep. The primary motivation for this computational perspective is not focussed on performance, but rather on providing easily-understandable, open-source software. This enables the presented results to be easily replicated and verified on any scientist's own machine, rather than requiring any specialised hardware or software setup. The simulations presented here are therefore necessarily primitive in comparison with the state-of-the-art \citep[e.g.][]{2005Natur.435..629S}, but provide examples that demonstrate what can be expected from such a toy model. Unless stated otherwise, all of the simulations use a total of 50,000 particles. I use standard N-body units, with $G=1.0$ and a total positive mass $M_{+}=1.0$. A softening parameter of $\epsilon=0.07$ was also used. Due to computational reasons, no matter creation was included in the current simulations.\footnote{While no matter creation is included in the simulations, this would only serve to add additional negative masses, which in most cases would only further strengthen the results.}

The simulations are presented in Sections~\ref{DMhalo-sim}, \ref{DMflatcurve}, and \ref{structure-sim}. In Section~\ref{DMhalo-sim}, I demonstrate that in the presence of negative mass material a positive mass galaxy naturally forms a dark matter halo with a radius that is several times larger than the positive mass component. Furthermore, due to the mutually-repulsive nature of negative masses, the formed halo is not cuspy. This provides a resolution of the cuspy-halo problem \citep[e.g.][]{2010AdAst2010E...5D}, and to my knowledge makes negative masses the only dark matter candidate that can provide a non-contrived solution. In Section~\ref{DMflatcurve}, I show that -- in the simulations -- the gravitational force exerted by this negative mass halo pushing inwards on the positive mass galaxy leads to a flat galaxy rotation curve. In Section~\ref{structure-sim}, I demonstrate that a mix of positive and negative mass material can result in structure formation. An initially uniform distribution of both positive and negative masses leads to the conventional filamentary and void-like structures observed in standard large-scale structure simulations. Moreover, the positive mass component of these simulations naturally becomes surrounded by negative mass material -- leading to ubiquitous dark matter haloes for every astrophysical object. In Section~\ref{runaway-sim}, I describe the constraints on runaway motion that are provided by the simulations.

   \begin{figure*}
   \centering
   \includegraphics[trim=0.0cm 0.0cm 0cm 0cm,clip=false,angle=0,origin=c,width=11.999cm]{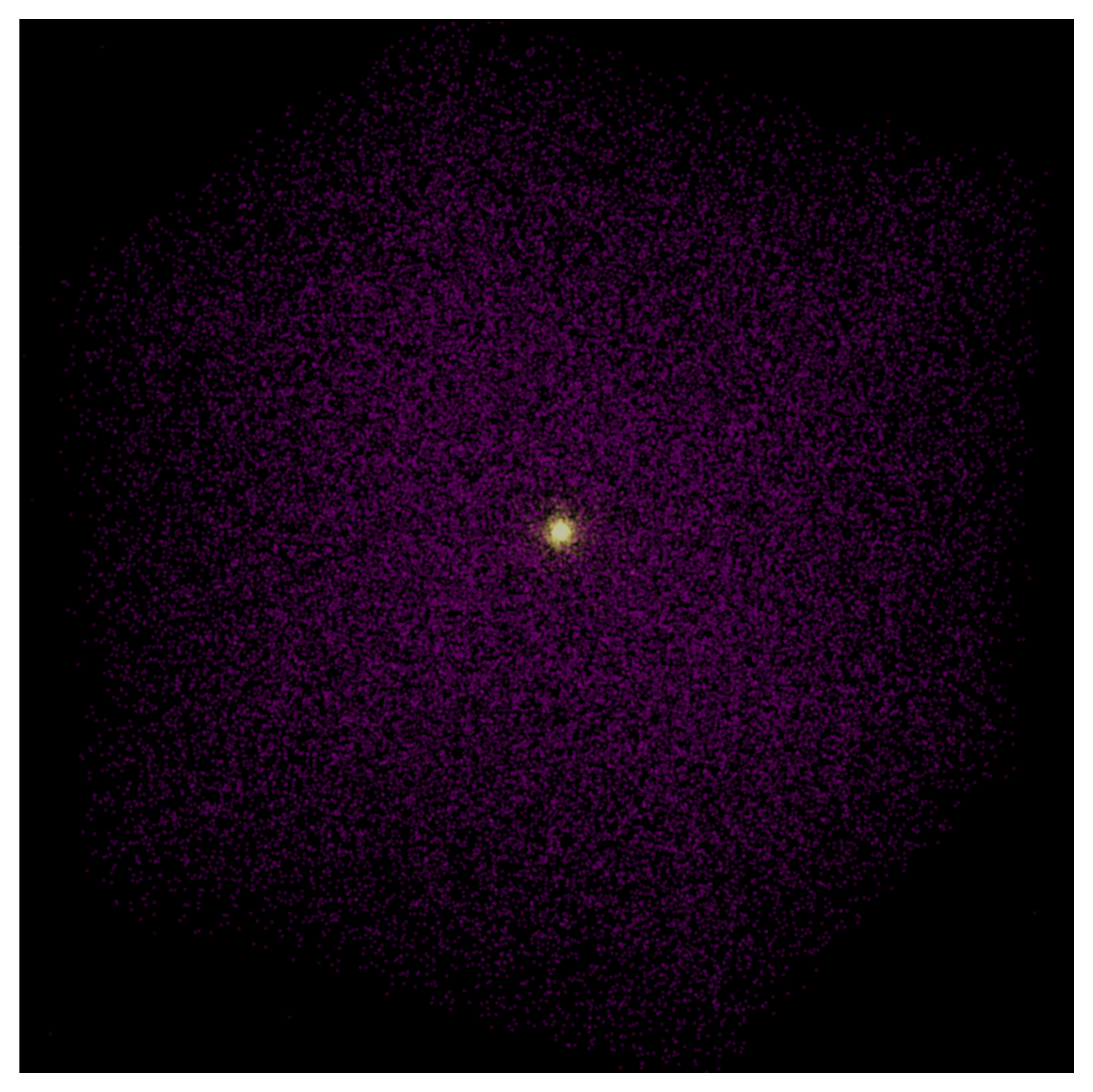}\\
      \vspace{-5pt}
      \includegraphics[trim=0.0cm 0.0cm 0cm 0cm,clip=false,angle=0,origin=c,width=11.999cm]{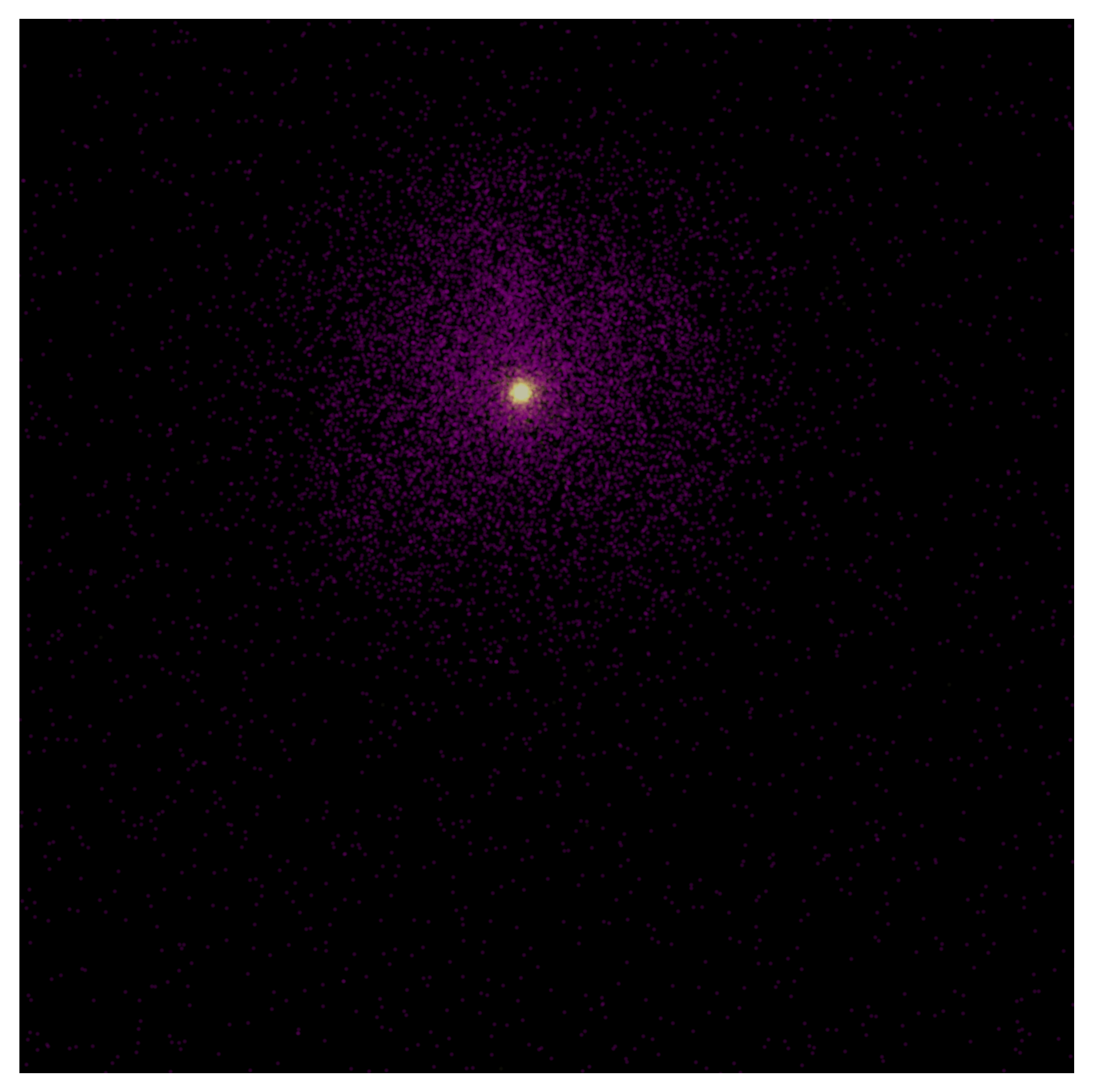}\\
   \caption{N-body simulations showing the formation of a non-cuspy dark matter halo from an initially motionless particle distribution of 45,000 negative masses (in purple), that surround a Hernquist-model galaxy of 5,000 positive masses (in yellow). Both the initial (top) and the final (bottom) time-steps are shown. An animated video from this simulation is available online.}
              \label{DMhalo}%
    \end{figure*}
    
\subsection{Formation of a dark matter halo}
\label{DMhalo-sim}
In the first set of simulations, a positive mass galaxy is located at the centre of the initial particle distribution. This positive mass galaxy is initialised with spherical-symmetry and following the conventional Hernquist model, with a scale radius equal to 1.0 \citep{1990ApJ...356..359H}. The Hernquist model is used to set the initial particle positions and velocities of the positive masses. The velocities consist of a radial component and are also built hot, with a velocity dispersion provided by a small Gaussian component with a standard deviation of 0.3. The galaxy comprises 5,000 particles, with a total positive mass of $M_{+}=1.0$. The initial positive mass particle distribution is located at the centre of a cube of negative masses with volume $200^3$. The initial conditions of these negative masses are set to be uniformly distributed in position and with zero initial velocity. The negative mass sea comprises 45,000 particles, with a total mass of $M_{-}=-3.0$. The simulation is scaled such that the positive mass galaxy has similar properties to the Milky Way, with a characteristic radius of 15~kpc and a mass of $5.8\times10^{11}M_{\odot}$. The simulation consequently runs over 21.5~Gyr with a timestep of 35.9~Myr. Each side of the box of negative masses has a length of 3~Mpc. Assuming spherical particles, each of the 45,000 negative masses therefore has an initial descriptive radius equivalent to 52~kpc between each particle.

The resulting particle distribution from this simulation is shown in Figure~\ref{DMhalo}. The full animated video from this simulation is available online. The negative masses at the outskirts of the cube are mutually-repelled by other surrounding negative masses and the cube begins to expand in volume, as discussed in Section~\ref{neg-cosmo}. Meanwhile, the negative masses within the central portion of the cube are attracted towards the positive mass galaxy. From their initially zero velocities, the negative mass particles are slushed to-and-fro from either side of the positive mass galaxy. Eventually, the negative mass particles reach dynamic equilibrium in a halo that surrounds the positive mass galaxy and which extends out to several galactic radii. The negative mass particles have naturally formed a dark matter halo.

\begin{figure}
   \centering
   \includegraphics[clip=false,angle=0,origin=c,width=\hsize]{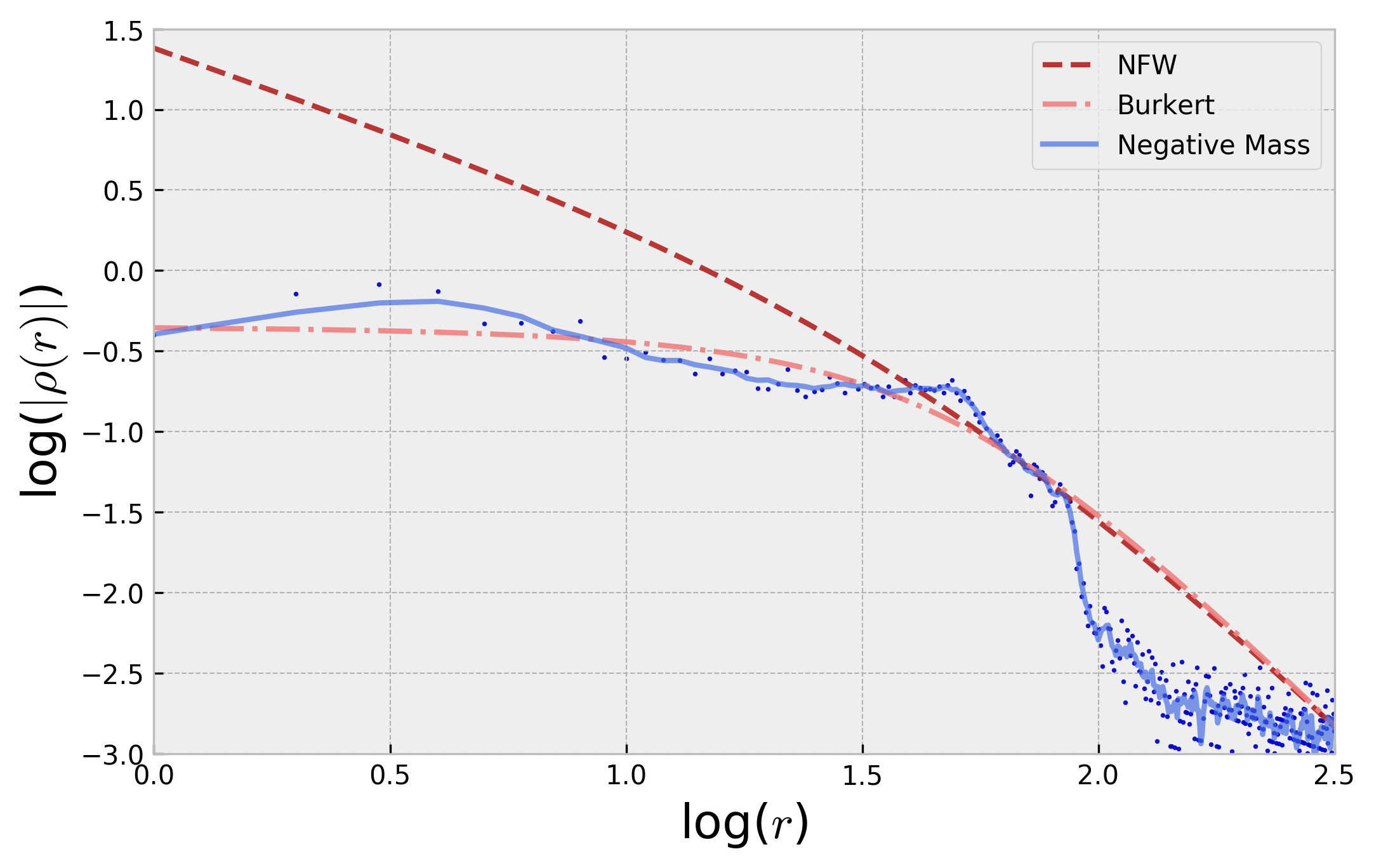}
   \caption{Plot of the magnitude of the density profile as a function of radius from the galactic centre, as extracted from the N-body simulations. The density profiles shown are: (i) as empirically determined for a negative mass cosmology from the N-body simulations (in blue), (ii) for a NFW profile (in dark red), and (iii) for an observationally motivated Burkert profile (in light red). The negative mass density profile is calculated in equally-spaced radial bins, with the measurements indicated by data points and overplotted with a moving average to guide the eye. The simulated dark matter halo is non-cuspy and best described by the Burkert profile. Beyond the radius of the negative mass halo, the radial profile of the diffuse negative mass background becomes visible. The sharp cut-off to the density profile at the edge of the halo may be further modified by matter creation. Negative masses can therefore simply reproduce the key features of observed dark matter density profiles in real galaxies.}
              \label{cuspyhalo}%
    \end{figure}

Observations of galaxies typically indicate an approximately constant dark matter density in the inner parts of galaxies, while conventional cosmological simulations of positive mass dark matter indicate a steep power-law-like behaviour. This is known as the core--cusp problem or the cuspy halo problem and is currently unsolved \citep{2010AdAst2010E...5D}. The negative mass halo that has formed in the simulations presented here can be clearly seen in the simulations to have a flat central dark matter distribution. The typically assumed positive mass particles that are used in conventional simulations are gravitationally attractive and thereby accumulate into a sharp cusp. However, negative mass particles are self-interacting and gravitationally repulsive -- thereby yielding a flat inner density profile. To demonstrate this, the extracted density profile from the simulations is shown in Figure~\ref{cuspyhalo}. It is worth noting that the precise form of the density profile will be further modified by matter creation. Nevertheless, the magnitude of the simulated negative mass density profile is shown and compared to both the cuspy Navarro--Frenk--White (NFW) profile derived from standard N-body simulations with positive mass dark matter \citep[e.g.][]{1997ApJ...490..493N}, and to the non-cuspy, observationally motivated, Burkert density profile \citep{1995ApJ...447L..25B}. The Burkert profile is clearly a much better representation of the simulated dark matter halo than the NFW profile. A negative mass cosmology can therefore provide a solution to the cuspy halo problem. This appears to be the only cosmological theory within the scientific literature that can explain and predict the distribution of dark matter in galaxies from first principles.

As the halo consists of negative mass particles, this will screen the positive mass galaxy from long-range gravitational interactions. This is similar to the screening effects that are seen in electrical plasmas with positive and negative charges, except this is a gravitational plasma with positive and negative masses. The negative mass sheath surrounding a typical galaxy effectively begins to shield the positive mass from external gravitational effects. The result is that in a negative mass dominated cosmology, positive mass galaxies would gravitationally attract each other and merge at a slower rate than in a conventional $\Lambda$CDM cosmology.

\subsection{Flattened rotation curves}
\label{DMflatcurve}
The rotation curves of galaxies have been shown to remain essentially flat out to several tens of kpc. This relation has been demonstrated using both H$\alpha$ observations \citep{1970ApJ...159..379R,1980ApJ...238..471R} and radio HI observations \citep{1981AJ.....86.1825B}. We can therefore attempt to measure the effect that negative masses have on the rotation curves of galaxies.

  \begin{figure}
   \centering
   \includegraphics[clip=false,angle=0,origin=c,width=\hsize]{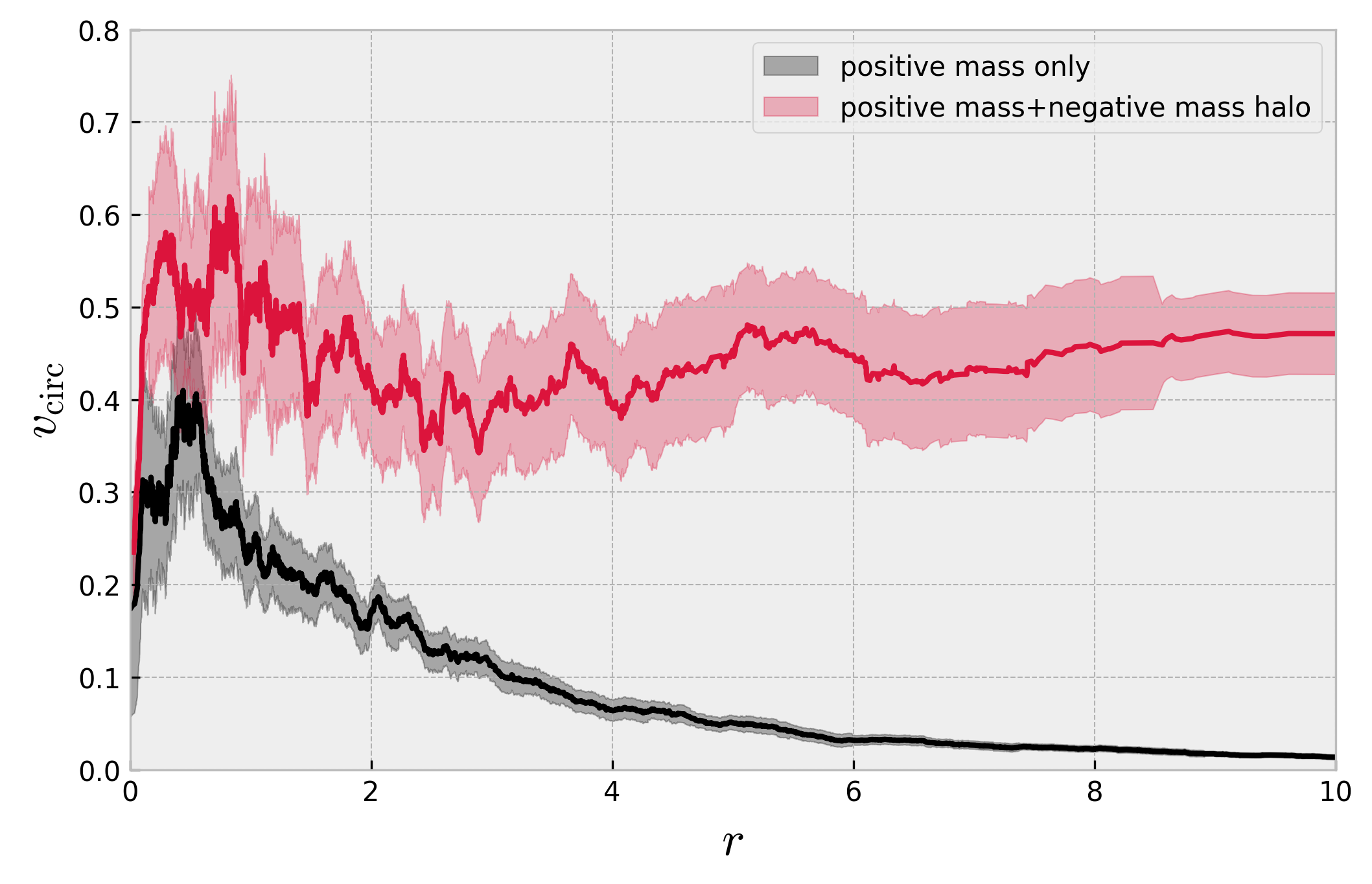}
   \caption{Plot of the circular velocity, $v_{\textrm{circ}}$, as a function of radius, $r$, as extracted from the N-body simulations. The rotation curves shown are: (i) for a positive mass galaxy consisting of 5,000 particles (in black), and (ii) for an identical positive mass galaxy with identical initial conditions, albeit now surrounded by an initially uniform distribution of 45,000 negative mass particles (in red). The rotation curves are plotted using a moving average for each set of 50 particles -- the means and standard errors in $v_{\textrm{circ}}$ for these particles are indicated by the solid lines and the shaded areas respectively. The rotation of the galaxy that consists of only positive mass clearly drops off following a Keplerian curve. Nevertheless, the galaxy with a negative mass halo has an essentially flat rotation curve that is slowly rising out to large galactic radii.}
              \label{rotationcurve}%
    \end{figure}
    
I use similar initial conditions as in Section~\ref{DMhalo-sim}. However, in order to reliably measure the rotation curve, the positive mass galaxy is setup as a kinematically cold system with no velocity dispersion and only a circular, orbital, velocity component. The rotation curve was first measured from a simulation with an initial particle distribution that consisted solely of 5,000 positive masses in a Hernquist model galaxy. The resulting rotation curve is indicated by the black line in Figure~\ref{rotationcurve}. The rotation curve for the positive mass galaxy clearly follows a Keplerian curve, with solid body rotation within the scale radius of the galaxy\footnote{A solid disk rotates such that the velocity increases linearly with radius.}, followed by a steady decline. The rotation curve was then also measured from another simulation with an identical particle distribution for the 5,000 positive masses, but now also surrounded by 45,000 initially uniformly distributed negative masses. The resulting rotation curve is indicated by the red line in Figure~\ref{rotationcurve}. The rotation curve for this positive mass galaxy with a negative mass halo also exhibits solid body rotation within the scale radius of the galaxy, but then appears to slowly increase, remaining essentially flat out to several galactic radii albeit with a slight positive incline. I emphasise that the only difference between these two simulations is that one contains only positive mass matter, whereas the other contains both positive mass matter and a negative mass halo. The negative masses have flattened the rotation curve of the galaxy. The inclusion of matter creation would be able to provide full dynamic equilibrium. Theoretical considerations that explain how negative masses can flatten the rotation curves of galaxies are detailed in Section~\ref{flatten-theory}.
    
\subsection{Structure formation}
\label{structure-sim}
In a universe filled with both positive and negative masses, one may raise the interesting question of whether the formation of large-scale structures could possibly take place. In standard structure formation simulations, a uniform distribution of positive mass particles is allowed to evolve. Small initial over- and under-densities in the particle distribution develop into filaments and voids similar to those seen in observations \citep[e.g.][]{2005Natur.435..629S}. The same process may therefore occur when negative masses are included. To test this hypothesis, simulations are setup within a cube-shaped region of volume $200^3$. Particles are located within this arbitrarily-sized region: 25,000 positive mass and a further 25,000 negative mass particles -- with the $x$, $y$, and $z$ coordinates of every particle being drawn from a uniform distribution. The total mass within the cube is $0.0$, with total positive and negative masses of $+1.0$ and $-1.0$ respectively. The initial velocities of all particles were equal to zero.  The simulation is scaled such that each side of the box has a length of 200~Mpc. The simulation runs over 21.5~Gyr with a timestep of 35.9~Myr. This provides a coarse mass resolution of $1.7\times10^{17}M_{\odot}\approx300,000$~Milky Ways per particle.

   \begin{figure*}
   \centering
   \includegraphics[trim=0.0cm 0.0cm 0cm 0cm,clip=false,angle=0,origin=c,width=11.999cm]{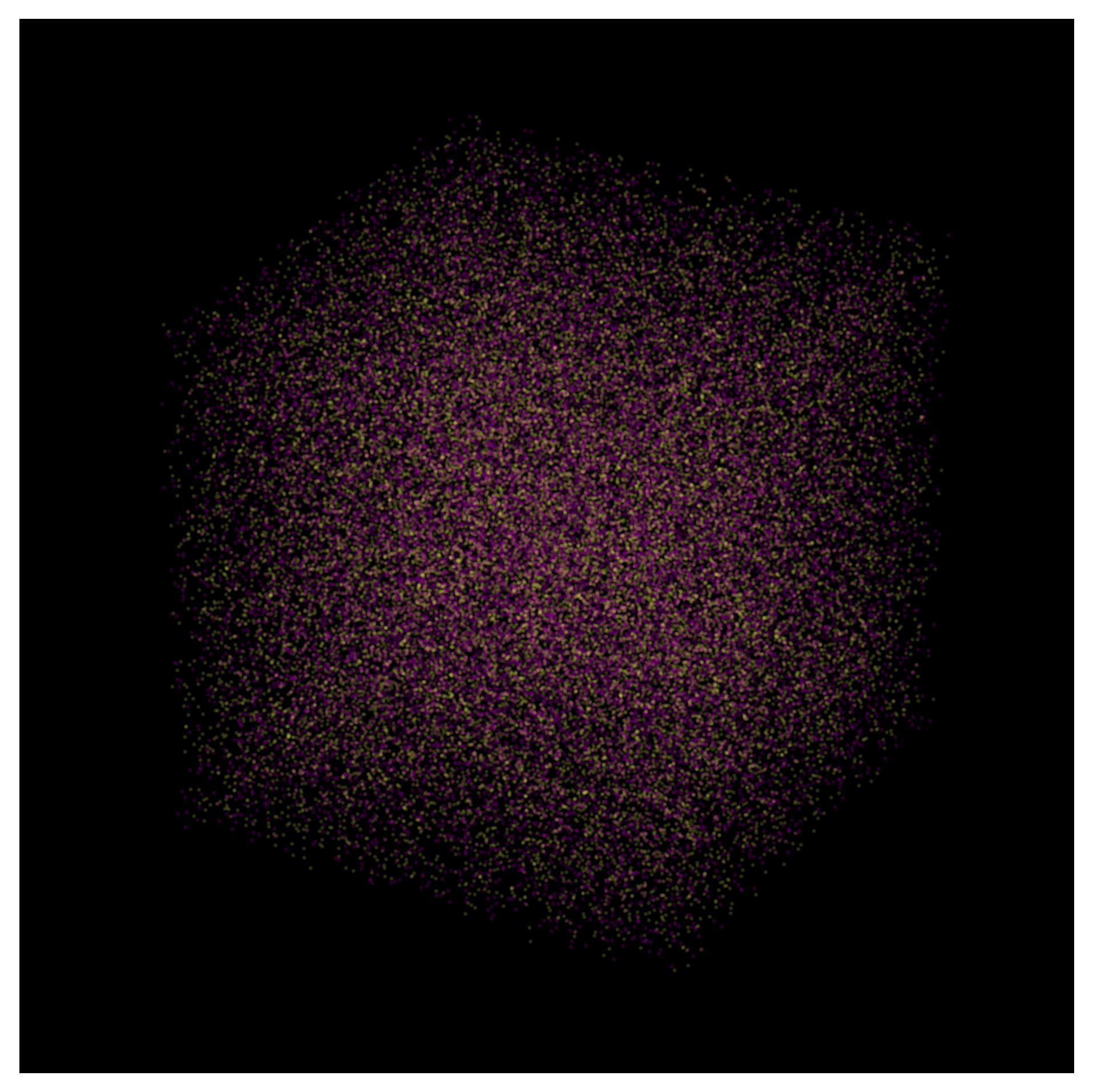}\\
      \vspace{-5pt}
      \includegraphics[trim=0.0cm 0.0cm 0cm 0cm,clip=false,angle=0,origin=c,width=11.999cm]{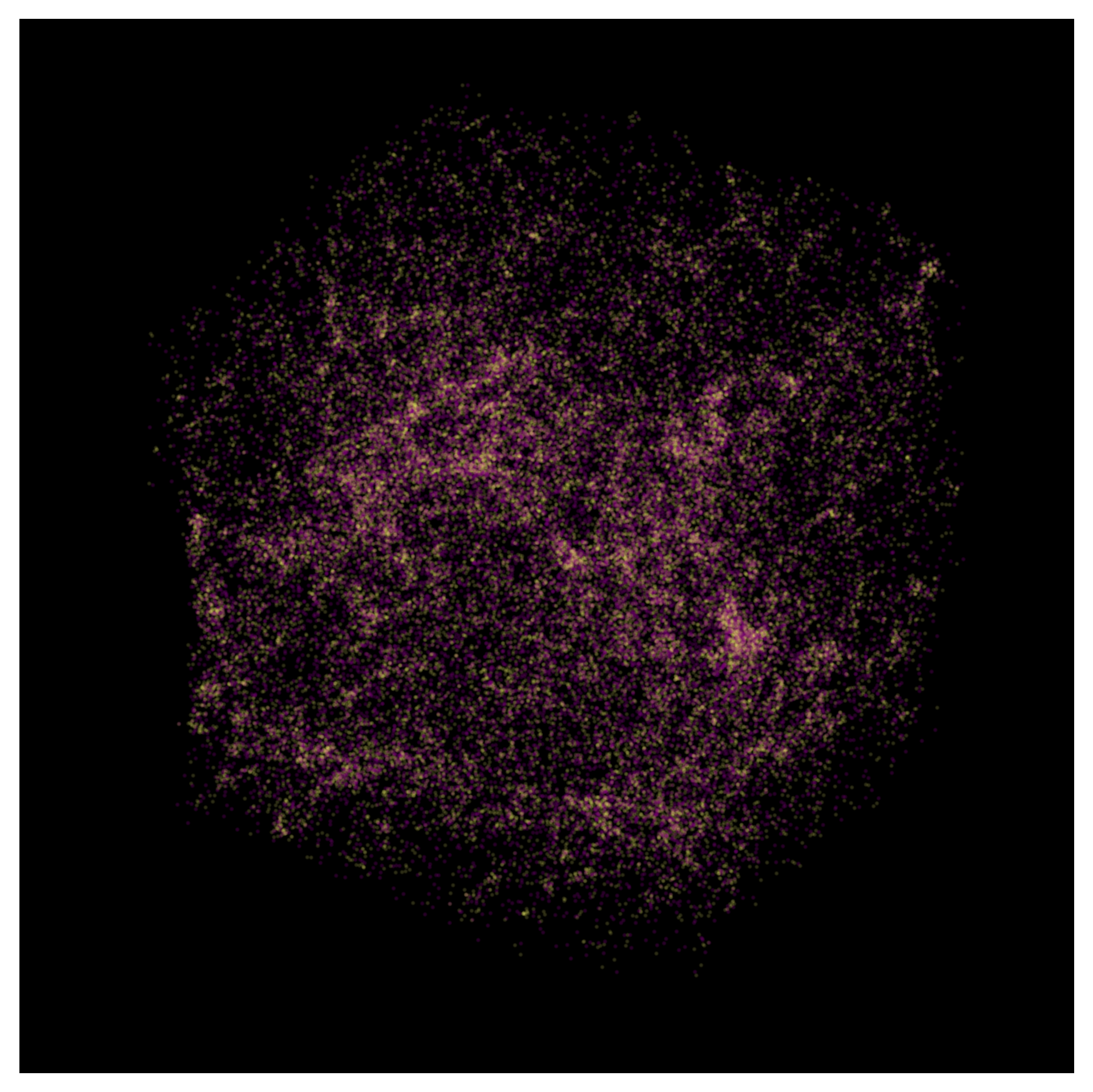}\\
   \caption{N-body simulations showing the formation of large-scale structure from an initially motionless, uniform, particle distribution of 25,000 positive masses (in yellow) and 25,000 negative masses (in purple). Both the initial (top) and the final (bottom) time-steps are shown. An animated video from this simulation is available online.}
              \label{structureformation-sim}%
    \end{figure*}
    
The resulting particle distribution from this simulation is shown in Figure~\ref{structureformation-sim}. The full animated video from this simulation is available online. As the simulation progresses, structure including filaments and voids clearly begin to be formed. Similarly to Section~\ref{DMhalo-sim}, the negative masses again naturally surround the positive masses, providing ubiquitous dark matter haloes. The initial particle distribution has formed a complex network that comprises filaments, voids, and rich clusters. From these early simulations, it is unclear which is the predominant effect: the additional pressure from the negative masses being attracted towards positive masses leading to more rapid structure formation than can occur in a positive mass only universe, or the mutual repulsion between negative masses tending to counteract this and leading to slower structure formation. Whichever the case, the presence of negative mass particles leads to modification of the relative spatial distribution between filaments and voids.\footnote{The consequences for structure formation could be substantially influenced by matter creation. Negative masses that are continuously created may feasibly have little influence on structures in the very early universe, as within the matter creation framework one could infer that few negative masses would exist at early-times.}

One-dimensional simulations of particles with a negative gravitational mass have also been reported in a recent study \citep{2018PhRvD..98b3514M}. More sophisticated N-body simulations with larger numbers of particles and more sophisticated initial conditions will be able to compare the resulting filaments and voids from these simulations with the observed large-scale structure in our Universe. Nevertheless, we have obtained a key result -- that structure formation is possible in a universe with negative mass.

\subsection{Runaway motion}
\label{runaway-sim}
In Section~\ref{intro}, we discussed the peculiar concept of runaway motion which can take place between positive--negative mass particle pairs, with such gravitational dipoles accelerating up to a speed equal to $c$. This has previously been perceived as a problem for theories of negative masses \citep[e.g.][]{bonnor1989}, partly due to the conventional maxim that massive particles cannot accelerate to $c$, and primarily due to the general reasoning that we do not observe such high-speed particles. Both components of this argument are however of dubious merit.

Firstly, the theory of positive--negative mass particle pairs provides clear rules that govern such interactions. The mechanics of these interactions are governed by the usual physical laws: the conservation of energy and momentum remain fundamental, and hence it is unclear why we should object to this potentially physical law of nature on grounds of aversion alone. Secondly, and more importantly, observations provide evidence for significant numbers of ultra-high-energy cosmic rays which are known to be extragalactic in origin, although the mechanism of their production remains a mystery \citep{pierreauger}. From this perspective, runaway motion is not a challenge for negative mass models, but is rather a useful observational constraint.

The idea that all negative masses in a universe should form gravitational dipoles and accelerate to high energies is not supported by the simulations presented here (which have a limited number of particles), in which no runaway particles can be identified. While runaway motion is a legitimate physical facet of negative mass particle interactions, the simulations indicate that this behaviour is only common for idealised particle pairs and occurs more rarely as a bulk behaviour within a negative mass fluid. This is likely as the particles in such a fluid are subject to numerous counteracting forces from the surrounding medium. One can assume that some amount of runaway particles must still exist, although these would likely be highly scattered by Brownian motion \citep[e.g.][]{landis1991}.

One possibility is that the softening parameter in our simulations, $\epsilon$, which affects short range interactions, could be preventing the formation of polarised mass dipoles that undergo runaway motion. To test this, the simulations were rerun with $\epsilon=0.0$. However, no runaway particles were detected in these simulations and the computational results remain unchanged. One can conclude that runaway motion must be sufficiently rare within a bulk fluid that the effect does not occur with any regularity in a simulation of 50,000 particles. Simulations with higher numbers of particles (of the order of millions) will be able to place numerical constraints on the runaway particle rate, in order to provide direct observational comparisons with ultra-high-energy cosmic ray detection rates.


\section{Compatibility with observations}
\label{compatibility}
At this stage, I have presented a toy model which predicts that the introduction of continuously-created negative masses to Einstein's field equations can behave in such a way as to resemble a cosmological constant. Furthermore, this model suggests that these negative masses can flatten the rotation curve of a galaxy. This suggests that negative masses could possibly be responsible for dark matter and dark energy. While this paper is primarily focussed on the theoretical and simulated consequences of such exotic matter, for the sake of completeness I now review the literature and consider any potential for compatibility between the toy model and contemporary cosmological observations.

\subsection{Supernovae observations of an accelerating Universe}
\label{supernovae}
In Section~\ref{neg-cosmo}, I found that there is a degeneracy between parameter estimates of $\Omega_{M-}$ and $\Omega_{\Lambda}$. In a conventional $\Lambda$CDM cosmology, where $\Omega_{M-}$ is taken to be zero, one could therefore infer a positive cosmological constant instead of a negative density parameter. I note that a model with non-zero $\Omega_{M-}$ and zero $\Omega_{\Lambda}$ would likely give a different expansion rate to one with $\Omega_{M-}$ and non-zero $\Omega_{\Lambda}$.

There is strong observational evidence from high-redshift supernovae that the expansion of the Universe is accelerating due to a positive cosmological constant \citep{1998AJ....116.1009R,1999ApJ...517..565P}. However, inspection of these results reveals that the observations themselves may demonstrate initial evidence for a negative mass dominated Universe. In both of these seminal works, the very reasonable assumption was made that all matter in the Universe has positive mass.

For the Bayesian fits in \citet{1999ApJ...517..565P}, the analysis assumed a prior probability distribution that has zero probability for $\Omega_{M}<0$. This former work notes that throughout the previous cosmology literature, completely unconstrained fits have generally been used that lead to confidence regions which include the part of the parameter space with negative values for $\Omega_{M}$. In other words, a probability of zero was assigned to a negative mass cosmology.

In \citet{1998AJ....116.1009R}, a working definition was applied such that negative values for the current deceleration (i.e.\ accelerations) were generated only by a positive cosmological constant and not from ``unphysical, negative mass density''. The entire supernovae analysis was re-run in this former work, in order to further test this point. The paper shows that demanding that $\Omega_{\Lambda} \equiv 0$, forces one to relax the requirement that $\Omega_{M}\ge0$ in order to locate a global minimum in their $\chi^2$ statistic. Upon relaxing their prior, they obtain ``unphysical'' values of $\Omega_{M}=-0.38\pm0.22$ and $\Omega_{M}=-0.52\pm0.20$ for their two different fitting approaches. This appears to indicate that the supernovae data are providing evidence for a negative density contribution, $\Omega_{M}=\Omega_{M+}+\Omega_{M-}<0$. This would be evidence for a negative mass dominated cosmology.

To assume that all mass in the Universe is positive is highly reasonable, as there has never been a pre-detection of such exotic material. However, as I have shown in this paper, negative mass density may not be unphysical. In fact, one can argue that its presence can be inferred from cosmological and galaxy rotation data, and it may possibly be able to provide an explanation for dark matter and dark energy. Independent and more contemporary analyses of supernovae have also continued to use positive-definite priors for $\Omega_{M}$, even with samples of up to 740 supernovae \citep{knop2003,shariff2016}. This further demonstrates that this is not a systematic that can be alleviated by better data, but rather a conceptual challenge with the data analysis.

I emphasise that by not constraining $\Omega_{M}>0$, we are also making one less assumption. One can argue that due to Occam's razor, a negative mass cosmology is the more parsimonious hypothesis. I therefore suggest that it is entirely plausible that the current observational data do not favour $\Lambda>0$, but rather are evidence for $\Omega_{M}<0$. As shown in Section~\ref{neg-cosmo}, when negative masses are continuously created, $\Omega_{M}<0$ can be equivalent to a negative cosmological constant.

One could argue that the aforementioned papers refer to the negative energy density of matter, which scales as $1/a^3$, while the negative energy density presented here is in the form of a cosmological constant. However, a cosmological constant is simply one form that can be taken by the negative energy density. The precise negative energy density is controlled by the parameter $\Gamma$, which can vary as a function of space and time, as detailed in Section~\ref{mattercreation-section}. Even in the case where $\Gamma$ does provide a cosmological constant, this affects only the observationally determined magnitude of the $\Omega_{M}$ measurements, and not the sign which remains negative. Although beyond the scope of this current paper, future works will be able to reanalyse the supernovae data and obtain updated measurements of the cosmological parameters when matter creation is fully included within the Bayesian analysis.

\subsection{CMB observations of a flat Universe}
\label{cmb-obs}
One of the major outcomes from measurements of the CMB has been locating the precise position of the first acoustic peak. This peak is on the degree scale, which implies that $k=0$ \citep[e.g.][]{2003ApJS..148..175S}. One could therefore suggest that this rules out the negative mass dominated cosmology, for which it has been shown that $k=-1$ (see Section~\ref{neg-cosmo}).

One simple explanation is that measurements of a flat universe from the CMB are just a local effect, with the $k=-1$ Universe being sufficiently large that it is not presently possible to detect any curvature. However, as shown in Section~\ref{AdS-space}, the negative mass dominated universe undergoes a cycle of expansion and contraction with a timescale of $\sqrt{-3\pi^2/\Lambda c^2}$. For a large universe in which the local geometry appears to be flat, it would therefore imply an especially low value for the magnitude of the cosmological constant -- otherwise the universe would have recollapsed before reaching such a size. In this case, a reanalysis of the CMB is not necessary, and the hypothesis of a $k=-1$ Universe could be considered compatible with existing observations. This is of course a possibility, but could also be perceived by a sceptical reader as a hand-waving way to allow any geometry for the Universe -- regardless of the observational data.

Alternatively, if we accept the observational evidence that the first CMB acoustic peak demonstrates that the Universe is flat within the confines of the $\Lambda$CDM model, we could instead consider the effects that negative masses would have on the location of the CMB peaks. Naively, in the open spatial geometry of the negative mass dominated cosmology, the position of the first CMB peak would be expected to be located at a considerably smaller angle. The ratio of angular distances for the negative mass universe and the conventional $\Lambda$CDM universe at redshift $z\sim1100$ (corresponding to the surface of last scattering) is
\begin{equation}
\left. \frac{d_{A}^{-}(a)}{d_{A}^{\Lambda\textrm{CDM}}(a)} \right\rvert_{z=1100} = 169 .
\label{ratio-cmb}
\end{equation}
Hence an astrophysical object in a negative mass universe at $z=1100$ would have an observed angular size 169 times less than in a $\Lambda$CDM cosmology. This is identical to the situation presented for a massless cosmology in \citet{2012A&A...537A..78B} (hereon BLC2012, also see Section~\ref{massless-cosmo}). I now extensively follow that earlier work. The angular position of the first CMB peak is defined by the angle under which the sound horizon is seen at recombination, which is given by
\begin{equation}
\theta = \frac{\chi_{s}(z_{\ast})}{d_{A}(z_{\ast})} ,
\label{theta-cmb}
\end{equation}
where $\chi_{s}(z_{\ast})$ is the sound horizon, $d_{A}(z_{\ast})$ is the angular distance, and $z_{\ast}$ is the redshift of the last scattering surface. The equivalent multipole is then given by $l_{a} \sim \pi/\theta$. By definition, the sound horizon is the distance that acoustic waves can propagate in a primordial plasma, which is typically assumed to only consist of positive masses. Accounting for the expanding universe, the distance of the sound horizon is given by 
\begin{equation}
\chi_{s} = \int_{0}^{t} c_{s} \frac{dt^{\prime}}{a(t^{\prime})} ,
\label{chi-cmb}
\end{equation}
where the speed of sound $c_{s} = c/\sqrt{3(1+R)}$, where $R$ is a corrective factor due to the presence of baryons \citep{1995ApJ...444..489H}. The value of $R$ is related to the baryon to photon ratio $\eta$ by $R=1.1\times10^{12}\eta/(1+z)$.\footnote{One may argue that tight-coupling of photons to baryons could erase all anisotropies, however such a hypothesis would critically assume a purely positive mass cosmology.}

The mechanism of sound generation in a negative mass and a $\Lambda$CDM cosmology differ drastically. In the standard $\Lambda$CDM cosmology, inhomogeneities are produced during the inflationary epoch. However, in a negative mass cosmology, sound waves would be generated at the interfaces between positive and negative mass dominated regions. BLC2012 show that the relevant time of interest for sound generation is therefore the epoch of the quark--gluon plasma transition which takes place at a temperature of $T\sim170$~MeV. I note that inflation is itself not required in this particular cosmology, which does not have an age or horizon problem (see Section~\ref{massless-cosmo}). An extended exposition can be found in BLC2012. The aforementioned paper shows that acoustic waves would propagate in the plasma while positive and negative masses are in contact. The expression for the angular position of the first CMB peak is then given by
\begin{equation}
l_{a} \sim \pi \frac{d_{A}}{\chi_{s}(z_{\ast})} .
\label{firstcmbpeak}
\end{equation}
Using this expression, BLC2012 obtain $l_{a}\sim160$ for the multipole of the acoustic scale. The standard measurement is $l_{a}\sim300$ \citep{2003ApJS..148..175S}. Rather than a discrepancy by a factor of $\approx 169$, there is substantial cancellation between the larger geometrical term (from the open spatial geometry of a negative mass universe) and the larger sound horizon (from the slow evolution of the expansion rate before recombination). Within a factor of approximately two, a negative mass universe can therefore predict the location of the first CMB peak and hence may be consistent with CMB observations. However, this model assumes that there is repulsion between positive and negative masses that leads to a subsequent gravitational decoupling, whereas the negative masses proposed in this paper obey the weak equivalence principle. In the latter case, sound can be continuously generated up until the present epoch. As this model contains numerous approximations, there appears to be the potential for a negative mass universe to be fully consistent with observations of the CMB. It is quite surprising in a negative mass cosmology with negative spatial curvature, that the first CMB acoustic peak can naturally emerge at the correct scale -- a simple back of the envelope approximation is already within a factor of two and consistent with the typically interpreted flat spatial geometry.

In an effort to identify any plausible mechanism that due to the nature of negative masses may allow for further adjustment of the CMB peaks, we can also speculate about various other compressive and expansive effects in the early Universe. In the standard $\Lambda$CDM model, sound waves can be generated from overdensities in the primordial plasma that originated from quantum fluctuations enlarged during inflation. These overdensities would gravitationally attract matter, while heat from photon--baryon interactions would seek thermal equilibrium and create an outward pressure. Counteracting gravity and pressure thereby give rise to oscillations analogous to sound waves. In a negative mass dominated universe, it is apparent that the sound generation mechanism would be modified. There would be two opposing effects: repulsive pressure from negative masses within an overdensity would tend to erase anisotropies, while conversely negative mass haloes would tend to surround positive mass baryons and increase both the effective gravitational attraction and the subsequent collapse of these overdensities. Matter creation would also exert a further influence. While these additional effects could modify the physics of sound generation in the early Universe and the predicted anisotropies in the CMB, the precise effects would depend upon the particle physics of the negative masses themselves, which is beyond the scope of this current paper.

In addition to this modification of CMB anisotropies, the sound waves used to derive $l_{a}\sim160$ are generated at the interfaces between positive and negative mass dominated regions. In these regions, the positive and negative masses would tend to interact. These interactions could lead to runaway motion and possible annihilation of positive--negative mass particle pairs, which would further affect the precise mechanism of sound generation. However, such effects are not considered here. Further consideration of these additional plausible physical effects, together with theoretical constraints for the second and third CMB peaks, can form a robust test to either validate or rule out the cosmic presence of negative mass.

In summary, while CMB modelling has provided an exceptional fit to observational data, the parameterisation of a model is only ever as good as the selected model itself. I therefore am suggesting that CMB physics has chosen a `correct' model -- a Universe with a cosmological constant and cold dark matter. However, it is possible that the true nature of dark matter and dark energy may have remained elusive due to the critical assumption that both of the $\Lambda$CDM components have positive energy. By allowing for energy to also be negative, it may be possible to show that this modified $\Lambda$CDM model can be fully compatible with the CMB. Nevertheless, I again emphasise that a reanalysis of the CMB is not essential for this purpose, as the Universe could simply be large with minimal local curvature.

\subsection{Galaxy cluster observations}
\label{gal-clusters}
Galaxy clusters have also played a significant role in establishing the standard $\Lambda$CDM model of cosmology \citep[e.g.][]{2011ARA&A..49..409A}. Such observations make a critical assumption -- that clusters are standard buckets that contain a representative mix of the constituent components of the cosmos. However, if the entirety of `empty' space is filled with negative masses that are continuously popping into existence, this would prevent clusters from being standard buckets and would suggest that they are intrinsically-biased towards positive mass regions of the Universe.

Within the presented toy model, galaxy clusters therefore do not represent standard buckets. Although not standard buckets, one could still anticipate that at least some observations of a few galaxies or galaxy clusters may have found hints of a negative mass. In fact, ``puzzling results'' in clusters such as negative masses have been discussed in the literature \citep{2005MNRAS.360..727A}. For example, Chandra observations of the merging cluster Abell 2034 found hints of a negative mass and therefore did not plot those data \citep{2003ApJ...593..291K}, regions of the mass profile in the galaxy NGC~4636 yielded ``unphysical'' negative masses \citep{2009ApJ...706..980J}, in the NGC~3411 galaxy group the total mass was found to decline with increasing radius -- requiring material with negative mass \citep{2007ApJ...658..299O}, measurements of galaxy clusters using the Sloan Digital Sky Survey yielded data that indicate a negative mass in poor clusters with fewer than five galaxies \citep{2005ApJ...633..122H}, a CMB cluster lensing study found a cluster with ``a fairly significant preference for negative mass'' \citep{2017arXiv170801360B}, and a number of strong and weak gravitational lensing studies have discussed or found indications of negative masses in reconstructed mass distributions \citep{2003MNRAS.345.1351E,2004ApJ...604..596C,2006A&A...451..395C,2006MNRAS.367.1209L,2007MNRAS.375..958D}. Perhaps these findings can be trivially explained by mundane observational biases and systematics. Nevertheless, given that we have identified other possible evidence for the influence of negative masses on other spatial scales, the repeated observation of negative mass in clusters appears to be yet another piece of evidence that allows us to infer the plausible existence of this exotic material.

\subsection{Overall compatibility with $\Lambda$CDM}
The current concordance cosmology is that of $\Lambda$CDM, which contains cold dark matter and a cosmological constant. In many respects, the negative mass dominated cosmology is a $\Lambda$CDM model -- with positive--positive mass interactions corresponding to baryons, positive--negative mass interactions corresponding to dark matter, and negative--negative mass interactions corresponding to dark energy. The standard $\Omega_{b}+\Omega_{\textrm{CDM}}+\Omega_{\Lambda}$ can therefore be reparameterised as $\Omega_{\textrm{++}}+\Omega_{\textrm{+-}}+\Omega_{\textrm{-}\textrm{-}}$, thereby providing a form of modified $\Lambda$CDM. It is not immediately clear whether the relative ratios of $4.9$\%, $26.8$\%, and $68.3$\% for $\Omega_{b}$, $\Omega_{\textrm{CDM}}$, and $\Omega_{\Lambda}$ respectively, would still hold upon a thorough observational reanalysis.

However, a key finding for the negative mass dominated cosmology is that the underlying universe is predicted to have $k=-1$, $\Omega_{M}<0$, and $\Lambda<0$. Claiming a cosmology with negative spatial curvature and negative cosmological constant would seem to be a heretical, renegade, and insane point of view. After all, the conventional $\Lambda$CDM cosmology (which is spatially flat with a positive cosmological constant, i.e.\ $k=0$, $\Omega_{M}>0$, and $\Lambda>0$) is based upon cutting-edge observational evidence derived from supernovae, the CMB, and galaxy clusters. While this is true, the interpretation of these observations has been derived using the critical assumption that all mass in the Universe is positive. While it is beyond the scope of this paper to attempt to fully recreate the entire impressive body of research into $\Lambda$CDM over the last 30 years, we have allowed ourselves to play devil's advocate and to have revisited the most key observational results, while also allowing mass to be negative. As scientists, we aim to be motivated purely by the scientific evidence alone and endeavour to remain entirely uninfluenced by confirmation bias. We have thus allowed ourselves to indulge in this unconventional thought experiment.

One can ask whether this negative mass cosmology could possibly be our cosmology. In Sections~\ref{supernovae} to \ref{gal-clusters}, I have shown that the cosmological parameters in the current concordance $\Lambda$CDM model can possibly be explained as a simple artefact that originates from the assumption that all matter in the Universe has positive mass. By allowing for negative masses within $\Lambda$CDM itself, I have surprisingly found that there is observational evidence that potentially supports and possibly even appears to favour a negative mass cosmology. I do not claim an all-encompassing or rigorous proof of a negative mass cosmology, but simply highlight that the toy model raises numerous interesting questions. Future work will be able to further test the compatibility with additional cosmological observations.


\section{Future considerations}
\label{discussion}
There are several outstanding theoretical challenges for a theory such as the one presented in this paper. I here provide some brief speculation as to the possibilities and future theoretical considerations.

\begin{itemize}
\renewcommand{\labelitemi}{$\bullet$}
\item It may be possible to directly validate this theory via the direct capture and detection of a negative mass particle. Particles undergoing runaway motion would be highly scattered due to Brownian motion (see Section~\ref{runaway-sim}), resulting in an observed isotropic distribution on the sky. At face-value, this is consistent with the origin of ultra-high-energy cosmic rays, and could lead to particles with energies above the Greisen--Zatsepin--Kuzmin (GZK) limit, such as the so-called Oh My God particle.
\item Although this paper only considers particles with identical inertial and gravitational mass, there are also a number of other negative mass models in which the inertial or gravitational mass alone may have an inverted sign. These models have recently been presented in \citet{2018PhRvD..98b3514M}, which provides one-dimensional structure formation simulations for these various scenarios. One such model is the Dirac-Milne universe \citep{2012A&A...537A..78B}, which explores the consequences of antimatter with negative gravitational mass (also see Section~\ref{massless-cosmo}). Experiments underway at CERN are expected to soon provide verification or refutation of these alternative negative mass models. It is possible that multiple forms of negative mass may possibly exist, and observational constraints will play an important role in testing these various scenarios.
\item It seems that the proposed negative mass fluid can be modelled as either matter or vacuum energy. It has previously been proposed that space-time arises as a form of large-scale condensate of more fundamental objects, that are typically of an unknown nature \citep[e.g.][]{2014PhRvL.112o1301L}. One could therefore speculate that the negative masses could be interpretable as a quantised form of energy associated with space-time itself.
\item The introduction of negative masses to the vacuum can also potentially provide a solution to the cosmological constant problem. The predicted vacuum energy can be a factor of $10^{123}$ larger than the observed value \citep[e.g.][]{2006gere.book.....H}. By invoking negative masses, the vacuum energy density can now take on essentially any value depending upon the precise cancellation of positive and negative energy states. If the negative oscillator modes exactly balance the positive modes, then $\rho_{\textrm{vac}}^{\textrm{quant}}=0$.
\item In theories of quantum gravity, gravitation is mediated by the graviton -- a massless, spin-2, boson. This means that any pair of negative masses would attract, and not repel as suggested in this theory. However, there are also theoretical arguments that gravitons cannot, and will not ever, be detected experimentally \citep{2006FoPh...36.1801R}. There appear to be two options: either it is possible that the graviton could be modelled as a bound state of a positive and a negative mass, in a theory of composite gravity or some other mechanism which provides a modification of graviton properties. Alternatively, this could also indicate that the proposed theory cannot be modelled by real, physical, particles, but rather by the presence of effective negative masses within a superseding theory.
\item Electrically-charged negative masses may tend to coalesce into highly charged clumps, eventually reaching a critical mass at which all other masses would be gravitationally repelled. This has previously been briefly described \citep{landis1991}. However, observational constraints on the abundance of negative mass compact objects indicate that such compact clumps cannot constitute a sizeable mass budget of the Universe \citep{2013ApJ...768L..16T}. One suggestion is that negative mass particles are always electrically neutral and remain in a diffuse form.
\item No attempt has been made to reconcile the presented theory with the standard model of particle physics. Can a viable Higgs mechanism allow for a negative mass? Is there a way to introduce negative masses into the standard model that could allow for the combination of fundamental forces at high energies, in a grand unified theory? Would supersymmetry be required? Is it possible that a negative mass particle travelling backwards in time may be measured as having a positive mass? These questions would be interesting future avenues that could be explored further by the particle physics community.
\end{itemize}


\section{Summary and conclusions}
\label{summary}
I have considered the introduction of negative masses and matter creation to cosmology, both via a theoretical approach and via computational simulations. Neither negative masses nor matter creation are new ideas. When considered individually, neither idea can explain modern astrophysical observations. This paper has reinvoked these two previous concepts and combined them together.

Commonly presumed issues with negative masses include incompatibility with general relativity (however, this was shown to be compatible in e.g.\ \citealt{bondi1957}), and the vacuum instability (which is not a bug, but rather a feature of the proposed theory, see Section~\ref{mattercreation-section}). By reintroducing the creation term into general relativity, but only for negative masses, it is possible to construct a toy model that has the potential to possibly explain both dark energy and dark matter within a simple and unified theoretical framework. Due to matter creation, a negative mass fluid can have $\omega=-1$. These hypothesised negative masses can push against positive mass galaxies and galaxy clusters, thereby modifying their dynamics. Under this theory, the cosmos contains a dynamic, motive, dark fluid, with dark matter and dark energy being modelled as the observed effects from positive mass matter `surfing' on this expanding fluid. As an illustrative concept, empty space-time would behave almost like popcorn -- with more negative masses continuously popping into existence.

From an astrophysical perspective, this cosmological theory surprisingly has some successes in describing observations. The derived cosmological model requires both negative spatial curvature, $k=-1$, and a negative cosmological constant, $\Lambda<0$. While such a proposal is a renegade and heretical one, it has been suggested that negative values for these parameters may possibly be consistent with cosmological observations, which have critically always made the reasonable assumption that mass can only be positive. When not making this extra assumption, Occam's razor indicates that the introduction of negative mass may possibly be a more parsimonious theory than the standard concordance $\Lambda$CDM model with $k=0$ and $\Lambda=0$. Considerable future work will be needed in order to fully explore the implications and prospects for this modified $\Lambda$CDM toy model.

The theory, simulations, and observations suggest that this particular cosmology has the following properties:
\begin{enumerate}
\item The geometry of the universe in this cosmology has negative spatial curvature, $k=-1$.
\item The continuous creation of negative masses can resemble a cosmological constant, with $\Lambda<0$.
\item Negative masses can give rise to a time-variable Hubble parameter.
\item Negative masses can be intrinsically attracted towards regions of positive mass, thereby leading to an increase in density that manifests itself as a dark matter halo that extends out to several galactic radii.
\item Due to mutual self-repulsion between negative masses, dark matter haloes formed from negative masses are not cuspy, and could thereby possibly provide a resolution of the cuspy-halo problem.
\item The rotation curves of galaxies can be flattened by the negative masses in the surrounding dark matter halo, however the curve is also predicted to increase linearly in the outermost regions of galaxies. This may be consistent with previous observational findings, which have found that most rotation curves are rising slowly even at the farthest measured point \citep[e.g.][]{1980ApJ...238..471R}.
\item Structure formation appears to be able to take place in a positive and negative mass universe, leading to the conventional suite of filaments, voids, rich clusters, and field galaxies.
\item Supernovae observations of a positive cosmological constant made the reasonable critical assumption that all mass is positive. Upon relaxing this assumption, the supernovae data of \citet{1998AJ....116.1009R,1999ApJ...517..565P} themselves derive a negative mass density in the Universe.
\item It appears that the first acoustic peak in the CMB could naturally emerge at the correct scale in a negative mass cosmology. Several additional physical effects on the CMB need to be fully considered, and determining the effects that negative masses have on the second, third, and higher order CMB peaks can enable a robust test to either validate or rule out the presence of negative mass in the Universe.
\item In this cosmology, negative masses are distributed throughout all of space-time, so that galaxy clusters cannot represent standard buckets. A number of galaxy cluster observations appear to have inferred the presence of negative mass in cluster environments.
\item The introduction of negative masses can lead to an Anti-de Sitter space. This would correspond to one of the most researched areas of string theory, the Anti-de Sitter/Conformal Field Theory correspondence, and if applicable to our own Universe, would suggest that string theory may possibly have direct physical applications.
\item Negative masses are predicted to produce a vacuum instability, which would suggest the vacuum itself is undergoing a slow and stable decay. In this cosmology, the universe would be taking on an increasingly negative energy state due to the continuous creation of negative masses. While such a vacuum instability is normally considered to be a theoretical insufficiency of negative masses, in this particular case it is not a bug, but rather a feature of the proposed cosmology.
\end{enumerate}

This accumulation of evidence could possibly indicate that while we cannot currently directly detect negative masses, we may have been able to infer the presence of these negative masses via their gravitational effects. These effects would seem bizarre, peculiar, and unfamiliar to us, as we reside in a positive mass dominated region of space. As the interactions between positive and negative masses are mediated by gravitation, the effects are typically fundamentally related to the physical scale -- generally requiring a sufficiently large accumulation of positive mass in order for negative masses to influence the dynamics of a physical system. One aspect that is particularly preposterous is the concept of runaway motion, but as quantum mechanics has shown, many absurd concepts constitute real, testable, and repeatable facets of nature.

There appears to be the potential and scope for this concept to be fully tested in order to make complete comparisons with observational data. A number of testable predictions have been made, including using cutting-edge telescopes such as the SKA, constraining the CMB acoustic peaks, and attempting direct detection from ultra-high-energy cosmic rays. Meanwhile, laboratory tests may be able to confirm whether antimatter could possibly be responsible for these gravitational effects -- although it would seem that a far more exotic material would likely be required. In addition, future state-of-the-art N-body simulations on GPUs with larger numbers of particles, that allow for the creation of negative masses, will help to provide a refined comparison with observations.

I here emphasise that several well-accepted theories can be modelled using non-real or effective negative masses. Air bubbles in water can be modelled as having a negative effective mass \citep{bubblebook}. For holes in semiconductor theory, electrons at the top of the valence band have a negative effective mass \citep{semiconductorbook}. The Casimir effect can be modelled using a region of negative energy density \citep{1988PhRvL..61.1446M}. Hawking radiation can be modelled using virtual negative mass particles that fall into the black hole \citep{hawking1975}. In the dark energy alternative of phantom energy (with $\omega<-1$), the excitations of the phantom field are negative mass particles \citep{2002PhLB..545...23C,2004AIPC..743...16C}. Even Bose--Einstein condensates have observable regions with negative effective mass \citep{2017PhRvL.118o5301K}. While the results in this paper appear to be consistent with vacuum states that have negative energy density, it is possible that these findings may imply a superseding theory that in some limit can be modelled by negative masses. In this way, the toy model could possibly be compatible with our own Universe, which may still satisfy the weak energy condition.

I suggest that a negative mass Universe is also a beautiful one. It naturally implies a symmetry, in which all physical systems are polarised into positive and negative states. A polarised cosmology that contains both positive and negative masses can literally bring balance to the Universe. This polarisation of the cosmos leads to a form of modified $\Lambda$CDM which seems to have the potential to quite possibly be able to describe our Universe in a more complete fashion than standard $\Lambda$CDM, with the distinct advantage that negative masses can offer a physical explanation for the natures of dark energy and dark matter. As it was Einstein that was the first to suggest that the cosmological constant could be modelled using negative masses, it seems that he potentially may have made two blunders. By not pursuing his own prediction, he may possibly have missed the chance to predict the existence of the mysterious dark aspects of our Universe. While it is certain that the negative masses discussed in this paper are gravitationally repulsive, it might be that the concept of negative mass is the most repulsive feature of all. Nevertheless, we should seriously consider the possibility that the perplexing nature of the dark Universe may feasibly have remained a mystery for 100 years due to a simple and pervasive sign error.


\begin{appendix} 
\section{Density evolution of a negative mass fluid in a non-expanding space}
\label{appendix-a}
I here consider the bulk properties of a negative mass fluid in a non-expanding space. To begin, I make a simplifying assumption and consider the effect of a cloud of massive particles in three dimensions, assuming spherical symmetry. The evolution of this massive fluid occurs in a dilute plasma, so that screening is unimportant and the particles interact via gravitational forces. I will consider the negative mass solutions for this `gravitational plasma'. While the negative mass solutions can be interpreted as the motion of negative mass particles, the solutions can also be considered as the flow of a negative mass fluid.

The majority of the matter in the Universe is known to exist as a low-density electrical plasma: solids, liquids and gases are uncommon away from planetary bodies. The cosmological principle therefore indicates that one can reasonably model the Universe as an isotropic, homogeneous, cloud of plasma. In this case, however, it is as a gravitational plasma, with particle interactions occurring between positive and negative masses rather than positive and negative electrical charges. I note that the expansion of space itself is not included in this unconventional plasma model of the Universe. In this Section, I extensively follow \citet{ivlev2013}, which provides similar solutions for a cloud of electrically charged particles that I will here modify for application to gravitating particles. I begin with a 1D planar problem and then develop analytical solutions for the 3D case. I neglect the pressure term in the equation of motion, assuming that thermal effects are negligible. Since the coordinate and density dependence of the particle mass, $M$, makes the problem non-linear (and likely intractable), I consider the case where $M$ is constant or an explicit function of time. These solutions therefore hold in cases where matter is constantly being created or annihilated.

\subsection{Planar 1D case}
I employ the Lagrangian mass coordinates $(s, t_{L})$ \citep{zeldovich-raizer2002}, where the Lagrangian time is $t_{L}=t$ and
\begin{equation}
s = \int_{0}^{x} dx^{\prime} n(x^\prime,t)
\label{coords}
\end{equation}
is the coordinate expressed via the local number density $n(x,t)$. I assume that the system remains symmetric with respect to $x=0$, this means that $v(0,t)=0$. The material and spatial derivatives can then be transformed from Eulerian coordinates using the rules
\begin{equation}
\frac{\partial}{\partial t} + v \frac{\partial}{\partial x} = \frac{\partial} {\partial t_{L}} ,
\end{equation}
and
\begin{equation}
\frac{\partial}{n \partial x} = \frac{\partial} {\partial s} .
\end{equation}

I now look at the resulting continuity and momentum equations for the density, $n$, and velocity, $v$, as well as Gauss' law for the self-consistent gravitational field, $\vec{g}$, produced by massive particles. For clarity, I omit the subscript $L$ for time, yielding
\begin{equation}
\frac{\partial n^{-1}}{\partial t} = \frac{\partial v} {\partial s} ,
\label{one}
\end{equation}
\begin{equation}
\frac{\partial v}{\partial t} + \nu v = \vec{g} ,
\label{two}
\end{equation}
\begin{equation}
\frac{\partial \vec{g}}{\partial s} = -4 \pi G M .
\label{three}
\end{equation}
Here $M$ is the mass of individual particles, and $\nu$ is the damping rate due to gas friction (i.e.\ kinematic viscosity). The number density of masses, $n$, is given by $N/V$ where $N$ is the total number and $V$ is the volume, or equivalently by $\rho/M$ where $\rho$ is the mass density.

I take the time derivative of Eq.~(\ref{one}), substitute $\partial^2 v/\partial s \partial t$ from Eq.~(\ref{two}), and use Eq.~(\ref{three}) to get
\begin{equation}
\frac{\partial^2 n^{-1}}{\partial t^2} +\nu \frac{\partial n^{-1}}{\partial t} = -4 \pi G M ,
\label{densityequation}
\end{equation}
which has a general solution
\begin{equation}
n^{-1}(s,t) = c_{1}(s) + c_{2}(s) e^{-\nu t} - \frac{4 \pi G M}{\nu}t ,
\label{solution1}
\end{equation}
where the constants $c_{1}$ and $c_{2}$ are to be determined from initial conditions. If the initial density is constant, $n(x,0)=n_{0}$ within the range $|x|\le x_{0}$, and so I obtain $n_{0}^{-1}=c_{1} + c_{2}$.

I now substitute $\partial n^{-1}/\partial t$ from Eq.~(\ref{solution1}) into Eq.~(\ref{one}) and integrate to obtain
\begin{equation}
v(s,t) = - \frac{4 \pi G M}{\nu}s - \nu e^{-\nu t} \int_{0}^{s} c_{2}(s^{\prime}) ds^{\prime} .
\label{solution2}
\end{equation}
If the particles were initially at rest, $v(s,0)=0$, then I obtain $c_{2}=-4 \pi G M/\nu^2$. Plugging this into Eq.~(\ref{solution1}) and using $n_{0}^{-1}=c_{1} + c_{2}$, I also obtain $c_{1}= n_{0}^{-1} +4 \pi G M/\nu^2$. The solutions of Eqs.~(\ref{one})--(\ref{three}) are therefore
\begin{equation}
\frac{n(t)}{n_{0}} = \left[1-\frac{\omega_{p}^2}{\nu^2} \left(\nu t - 1 + e^{-\nu t} \right) \right]^{-1} ,
\label{1Dresult-1}
\end{equation}
\begin{equation}
v(s,t) = -\frac{\omega_{p}^2}{\nu} \frac{s}{n_{0}} \left(1 - e^{-\nu t} \right) ,
\label{1Dresult-2}
\end{equation}
\begin{equation}
\vec{g} = -4 \pi G M s ,
\label{1Dresult-3}
\end{equation}
where $\omega_{p} = \sqrt{4 \pi G M n_{0}}$ is the initial plasma frequency of the massive cloud. These solutions are also valid when $M$ is an explicit function of time, for example due to matter creation or matter annihilation. I have therefore now obtained general solutions for (i) the local number density, (ii) the flow velocity, and (iii) the gravitational field. Eqs.~(\ref{1Dresult-1})--(\ref{1Dresult-3}) are all described as functions of the Lagrangian mass coordinates, ($s$, $t$). I note that via this mathematical representation of fluid flow we are tracking the locations of individual fluid particles, rather than using fixed-space Eulerian coordinates. The use of Lagrangian coordinates allows us to consider the fluid from a more cosmological perspective, since we are interested in the history of individual particles as a function of time.

We first discuss the local number density as a function of time, by considering an asymptotic analysis of Eq.~(\ref{1Dresult-1}). At early-times, $\nu t \ll 1$, and using a Maclaurin series expansion for $e^{-\nu t}$ yields
\begin{equation}
\frac{n(t)}{n_{0}} \approx \left[1-\frac{\omega_{p}^2 t^2}{2} \right]^{-1} = \left[1-\frac{4 \pi G M n_{0} t^2}{2} \right]^{-1} \textrm{(for }\nu t \ll 1\textrm{)}.
\label{density-earlytimes}
\end{equation}
At later-times, $\nu t \gg 1$, and $e^{-\nu t} \approx 0$ which leads to
\begin{equation}
\frac{n(t)}{n_{0}} \approx \left[1-\frac{\omega_{p}^2 t}{\nu} \right]^{-1} = \left[1-\frac{4 \pi G M n_{0} t}{\nu} \right]^{-1} \textrm{(for }\nu t \gg 1\textrm{)}.
\label{density-latetimes}
\end{equation}
Consequently, at $\nu t \sim 1$, the time-scaling switches from $n \propto t^{-2}$ ($\nu t \ll 1$) to $n \propto t^{-1}$ ($\nu t \gg 1$). I note that this does not consider matter creation or annihilation. A more general solution could be considered where either the annihilation of positive masses or the creation of negative masses leads to a constant density evolution with time. Assuming no matter creation or annihilation, the evolution of Eqs.~(\ref{density-earlytimes}) and (\ref{density-latetimes}) are dependent on whether $M$ is positive (in which case the density increases with time) or negative (in which case the density decreases with time).

In the case where $M>0$, the gravitational plasma of positive masses collapses through mutual gravitational attraction -- with the density increasing over time. However, the density cannot increase without bound. As the fluid becomes sufficiently dense, the positive mass particles will begin to interact on a scale equivalent to an effective particle radius. Furthermore, the plasma will eventually cease to be sufficiently dilute that the screening is unimportant and particles will begin to experience short-range interparticle interactions. The density will then evolve following some other form, of which we are not here concerned. Although I highlight that this is a 1D treatment in a non-expanding space, this solution describes the gravitational evolution of a positive mass cloud.

In the case where $M<0$, the gravitational plasma of negative masses now undergoes a gravitational explosion -- with the fluid becoming more dilute and the density decreasing over time. Such a cloud would expand persistently due to gravitational repulsion. Although this is also a 1D treatment in a non-expanding space, this solution describes the gravitational expansion of a negative mass cloud. In the case of no matter creation, we can see that the expansion is characterised by a uniform stretching, with the density being constant in space and decreasing monotonically with time. Consequently, Eq.~(\ref{coords}) is reduced to a simple relation $s=n(t) x$ and the cloud boundary $x_{b}(>0)$ is determined by the condition $s_{b}=n(t) x_{b}(t) = n_{0} x_{0}$. However, a more general solution that includes matter creation can also allow the density of negative masses to remain constant as a function of time, and can be trivially parameterised by treating $M$ as $M(t)$ and equivalently $\rho=n(t) M(t) = \textrm{constant}$ in Eqs.~(\ref{density-earlytimes}) and (\ref{density-latetimes}). In this case of matter creation, the expansion is still characterised by a uniform stretching, and if matter is created at a sufficient rate, the density can remain constant in both space and time.

I now discuss the flow velocity, $v$, as a function of time, by considering an asymptotic analysis of Eq.~(\ref{1Dresult-2}) when $M<0$. It can be determined that $v$ increases linearly with $x$ and attains a maximum at $x_{b}$. When $\nu t \gg 1$, the maximum velocity at the boundary tends to a constant value of $v_{b} = x_{0} \omega^2_{p}/\nu$, and corresponds to a balance of the gravitational and frictional forces. As $\nu$ tends towards zero, I obtain
\begin{equation}
v_{b} = \lim_{\nu \to 0} -\frac{\omega_{p}^2}{\nu} x_{0} \left(1 - e^{-\nu t} \right) = \omega_{p}^2 x_{0} t,
\end{equation}
which shows that in the absence of friction, $v$ grows linearly with time at the boundary.

One could ask what fate awaits overdensities and structure that may exist within this expanding negative mass fluid. I note that the functions $c_{1}(s)$ and $c_{2}(s)$ in Eq.~(\ref{solution1}) can also be derived for a more general case, when the initial density $n_{0}(x)$ is not constant, or the initial velocity $v_{0}(x)$ is not zero. This leads to $c_{1}(s) + c_{2}(s) = n_{0}^{-1}(s)$ and $c_{2}(s) = -4 \pi G M/\nu^2 - v_{0}^{\prime}(s)/\nu$, where $n_{0}(s)$ and $v^{\prime}_{0}(s) \equiv dv_{0}(s)/ds$ can be obtained from $n_{0}(x)$ and $v_{0}(x)$ respectively, by employing the relation $x(s)$ from Eq.~(\ref{coords}). Consequently, in the limit $\nu t \gg 1$, the solutions for $n$ and $v$ do not depend on the initial conditions and tend to the behaviour of Eqs.~(\ref{1Dresult-1}) and (\ref{1Dresult-2}) at later-times.

\subsection{Spherically symmetric 3D case}
I have shown that the solution for a 1D cloud has distinct characteristic features. In homogeneous steady initial conditions, the density of the expanding cloud remains homogeneous and the velocity increases linearly towards the cloud boundary. This is asymptotically true for arbitrary initial conditions. One could therefore hypothesise that these characteristics are independent of dimensionality and will hold in a spherically symmetric 3D cloud.

To test this hypothesis, I continue to follow \citet{ivlev2013} and write the analogous 3D equations to Eqs.~(\ref{one})--(\ref{three}) in Eulerian coordinates, which yields
\begin{equation}
\frac{\partial n}{\partial t} + \frac{1}{r^2} \frac{\partial (r^2 n v)}{\partial r} = 0 ,
\label{one3D}
\end{equation}
\begin{equation}
\frac{\partial v}{\partial t} + v \frac{\partial v}{\partial r} + \nu v = \vec{g} ,
\label{two3D}
\end{equation}
\begin{equation}
\frac{1}{r^2} \frac{\partial (r^2 \vec{g})}{\partial r} = -4 \pi G M n.
\label{three3D}
\end{equation}
I use the following starting point for the solutions to $n$, $v$, and $\vec{g}$:
\begin{equation}
\frac{n(t)}{n_{0}} = \left( \frac{r_{0}}{r_{b}(t)} \right)^3 ,
\label{3Dresult-1}
\end{equation}
\begin{equation}
v(r,t) = \frac{r}{r_{b}(t)} \dot{r}_{b}(t) ,
\label{3Dresult-2}
\end{equation}
\begin{equation}
\vec{g}(r,t) = -\frac{4 \pi}{3} G M n(t) r ,
\label{3Dresult-3}
\end{equation}
with $r \le r_{b}$ (for simplicity, I again consider a homogeneous initial distribution). These initial solutions ensure that Eqs.~(\ref{one3D})--(\ref{three3D}) are satisfied identically.

By substituting Eq.~(\ref{3Dresult-2}) into Eq.~(\ref{two3D}), then substituting the solutions for $\vec{g}$ from Eq.~(\ref{3Dresult-3}) and $n(t)$ from Eq.~(\ref{3Dresult-1}), I obtain
\begin{equation}
\ddot{r}_{b}(t) + \nu \dot{r}_{b}(t) + \frac{\omega_{p}^2}{3} \frac{r_{0}^3}{r_{b}(t)^2}= 0 .
\end{equation}
I now define, $N(t) = r_{b}(t)/r_{0}$, $\dot{N}(t) = \dot{r}_{b}(t)/r_{0}$, and $\ddot{N}(t) = \ddot{r}_{b}(t)/r_{0}$, and derive
\begin{equation}
\ddot{N} + \nu \dot{N} + \frac{\omega_{p}^2}{3} N^{-2}= 0 ,
\label{equation23}
\end{equation}
which can be more generally written as
\begin{equation}
\ddot{N} + \nu \dot{N} + \frac{\omega_{p}^2}{D} N^{-D+1}= 0 .
\label{main3Dequation}
\end{equation}
where the dimensionality of the space is represented by $D = 1,2,3$. Similarly from Eq.~(\ref{3Dresult-1}), I can more generally write that $N(t) = r_{b}(t)/r_{0} \equiv \left( \frac{n(t)}{n_{0}} \right)^{-1/D}$. Consequently, when $D=1$ then $N(t)=\left( \frac{n(t)}{n_{0}} \right)^{-1}$. Subsequently, Eq.~(\ref{main3Dequation}) reduces to Eq.~(\ref{densityequation}) and I obtain the 1D solutions from Eqs.~(\ref{1Dresult-1}) to (\ref{1Dresult-3}). When $D=2,3$, then one cannot solve Eq.~(\ref{main3Dequation}) analytically. However, some special cases can be considered via an asymptotic analysis.

Firstly, I can consider Eq.~(\ref{equation23}) in the case when $\nu t \gg 1$, such that $\ddot{N}$ is negligible. This yields
\begin{equation}
\frac{n(t)}{n_{0}} \approx \left[3 k_{1}-\frac{\omega_{p}^2 t}{\nu} \right]^{-1},
\end{equation}
which as the initial density $n(0)=n_{0}$, at $t=0$, this leads to $k_{1}=1/3$ which provides an identical solution to Eq.~(\ref{density-latetimes}) for the 1D case. Hence, I can conclude that asymptotically, the density evolution does not depend on the dimensionality and that our initial result holds when $D=3$. Hence for any case where $\nu\neq0$ -- and independent of the initial conditions, the density decays asymptotically as $n\propto t^{-1}$.

Secondly, I can consider Eq.~(\ref{equation23}) in another special case, when $\nu=0$, so that the equation can be directly integrated. In this case, the first integral is $\dot{N}^2=\frac{2}{3} \omega_{p}^2 (1-N^{-1})$, which has the solution
\vspace{-3pt}
\begin{equation}
\sqrt{N(N-1)}+\ln{\left(\sqrt{N}+\sqrt{N-1}\right)} = \sqrt{\frac{2}{3}}\omega_{p}t
\end{equation}
such that $N(t)=\left( \frac{n(t)}{n_{0}} \right)^{-1/3}$. The time evolution at small $t$ is then similar to Eq.~(\ref{density-earlytimes}), while the evolution at large $t$ is given by $n \propto t^{-3}$. I note that this 3D frictionless case is likely the most relevant for a cloud of negative masses, which should have low friction due to their mutual repulsion. We therefore can consider the density evolution of a perfect fluid of negative masses, with no viscosity, to be characterised by $n \propto t^{-3}$.

In conclusion, the expansion of a cloud/fluid of negative mass particles is described by analytical solutions. In the case of no matter creation, we can see that the expansion is characterised by a uniform stretching, with the density being constant in space and decreasing monotonically with time. In the case where matter is being created, the expansion is still characterised by a uniform stretching, but if matter is created at a sufficient rate then the density can remain constant in both space and time. Irrespective of whether matter creation is taking place, the density distribution remains homogeneous across the cloud, and the velocity increases linearly towards the cloud boundary. I note that this exercise is characterised within a non-expanding space and therefore provides the local properties for a cloud of negative masses, rather than full cosmological solutions. The precise analytical solutions for the expansion of this fluid would be modified in proximity to positive masses. In particular, such a gravitational plasma will begin to exhibit screening effects similar to those observed in an electrical plasma. It has been suggested that negative mass screening effects would have the ability to attenuate gravitational waves, with the plasma being opaque to frequencies below the plasma frequency \citep{2014PhRvD..90j1502M}.
\end{appendix}


\begin{acknowledgements}
I am extremely grateful to my dear wife, Kristina, for all her kind advice in drafting this document and for enduring many boring monologues about negative masses over a long time span. I also thank the anonymous referee for thorough comments, which helped to substantially improve the paper. I would also like to thank Takuya Akahori for helpful comments on structure formation in a negative mass filled Universe that inspired further simulations, and both Wes Armour and Will Potter for kindly providing constructive and insightful feedback on a draft of this paper. Any mistakes in this hopefully interesting paper are entirely my own.
\end{acknowledgements}


\end{document}